# Evidence for Pseudogap Phase in Cerium Superhydrides: $CeH_{10}$ and $CeH_9$

Dmitrii Semenok[1], Jianning Guo[2], Di Zhou[1, *], Wuhao Chen[3,4], Toni Helm[5], Alexander Kvashnin[6], Andrei Sadakov[7], Oleg Sobolevsky[7], Vladimir Pudalov[7,8], Chuanying Xi[9], Xiaoli Huang[2, *], and Ivan Troyan[10, *]

[1] Center for High Pressure Science & Technology Advanced Research, Bldg. #8E, ZPark, 10 Xibeiwang East Rd, Haidian District, Beijing, 100193, China

[2] State Key Laboratory of Superhard Materials, College of Physics, Jilin University, Changchun 130012, China

[3] Quantum Science Center of Guangdong–Hong Kong–Macao Greater Bay Area (Guangdong), Shenzhen, China

[4] Department of Physics, Southern University of Science and Technology, Shenzhen 518055, China

[5] Hochfeld-Magnetlabor Dresden (HLD-EMFL) and Würzburg-Dresden Cluster of Excellence, Helmholtz-Zentrum Dresden-Rossendorf (HZDR), Dresden 01328, Germany

[6] Skolkovo Institute of Science and Technology, Skolkovo Innovation Center, Bolshoy Boulevard, 30/1, Moscow 121205, Russia

[7] V. L. Ginzburg Center for High-Temperature Superconductivity and Quantum Materials, P. N. Lebedev Physical Institute, Russian Academy of Sciences, Moscow 119991, Russia

[8] National Research University Higher School of Economics, Moscow 101000, Russia

[9] Anhui Province Key Laboratory of Condensed Matter Physics at Extreme Conditions, High Magnetic Field Laboratory of the Chinese Academy of Science, Hefei 230031, Anhui, China.

[10] Shubnikov Institute of Crystallography, Federal Scientific Research Center Crystallography and Photonics, Russian Academy of Sciences, 59 Leninsky Prospekt, Moscow 119333, Russia

Corresponding authors: Di Zhou (di.zhou@hpstar.ac.cn), Xiaoli Huang (huangxiaoli@jlu.edu.cn) and Ivan Troyan (itrojan@mail.ru)

**Abstract**

Polyhydride superconductors have been shown to possess metallic properties with a Bardeen-Cooper-Schrieffer-type superconducting ground state. Here, we provide evidence for unconventional transport associated with a pseudogap phase in cubic cerium superhydride $CeH_{10}$ ($T_C = 116$ K) at a pressure of 115-125 GPa. A large negative magnetoresistance in the non-superconducting state below 90 K, quasi $T$-linear electrical resistance, and a sign-change of its temperature dependence mark the emergence of this phase. We studied the magnetic phase diagrams and the upper critical fields $B_{C2}(T)$ of $CeH_{10}$, $CeH_9$, and $CeD_9$ in pulsed fields up to 70 T. $B_{C2}(T)$ of $CeH_9$ and $CeD_9$ exhibits pronounced saturation at low temperatures in accordance with the Werthamer-Helfand-Hohenberg model, whereas $CeH_{10}$ stands out in particular, as it does not obey this model. Our observations, therefore, reveal the unconventional nature of the non-superconducting state of cerium superhydride $CeH_{10}$.

**Introduction**

The family of compressed polyhydrides has been found to exhibit high-temperature superconductivity in the megabar regime [1-5]. For the cerium hydrides $CeH_9$ and $CeH_{10}$ relatively high critical superconducting temperatures ($T_C$) up to 120 K have been observed at pressures of about 1 Mbar [6-7], rendering them a convenient platform for in-depth studies of hydride superconductivity (SC). This pressure is well within reach for modern high-pressure techniques. It, therefore, enables experiments on decently sized samples with dimensions of about 0.1 mm with a reduced risk of diamond anvil cell (DAC) failure during the laser heating synthesis. Many polyhydrides, such as $LaH_{10}$ [1], $YH_6$ [2-3], $YH_9$ [3], $YH_4$ [4], $H_3S$ [5], have rather high upper critical magnetic fields $B_{C2}(0)$ around 70-140 T or above, which poses a challenge for experiments. By contrast, the cerium superhydrides possess exceptionally low $B_{C2}(0)$ values ranging between 25 and 35 T [7]. This allows for comprehensive investigations of their magnetic phase diagram and, in particular, the normal state beyond $B_{C2}$ down to lowest temperatures.

So far, such studies with hydrides have only been performed for a few low-$T_C$ superconductors (e.g. YH$_4$ [8], SnH$_4$ [9]). For instance, tin tetrahydride SnH$_4$ exhibits not only a linear $B_{C2}(T)$, but also a $B,T$-linear magnetoresistance over a wide range of temperatures and magnetic fields. The linear temperature dependence of $B_{C2}$ contradicts the predictions of the Werthamer–Helfand–Hohenberg model [10-12]. This model is based on the Bardeen-Cooper-Schrieffer (BCS) theory of superconductivity and the anomalous behavior of the upper critical magnetic field may point to an unconventional pairing mechanism in some compressed polyhydrides.

In our work, we performed a comprehensive study of the magnetic phase diagram by means of magnetoresistance and Hall-effect measurements of CeH$_{10}$, CeH$_9$, and deuteride CeD$_9$ synthesized by laser heating from metallic cerium and ammonia borane (NH$_3$BH$_3$ or ND$_3$BD$_3$) at 112-130 GPa. Most importantly, we conduct an extended study of the normal-state transport response in pulsed magnetic field up to 70 T. Our observations in CeH$_{10}$ provide strong evidence for a distinct region in its phase diagram, where a pronounced negative magnetoresistance emerges. This suggests the existence of a pseudogap phase, similar to the cuprate superconductors, which is further supported by its unconventional temperature dependence.

**Results**

*Cerium superhydrides CeH$_9$ and CeH$_{10}$*

Three diamond anvil cells (DACs) I1, H1, D1 were loaded with metallic cerium (99.99%) and ammonia borane (NH$_3$BH$_3$, DACs I1 and H1) or its deuterated analogue, ND$_3$BD$_3$ (DAC D1), obtained by the procedure described in Refs. [2,7]. BeCu (H1, D1) and NiCrAl (I1) non-magnetic diamond anvil cells with anvils culet diameter of 100 μm were used in our experiments. A sputtered four-probe Au/Ta (DAC I1) or Mo (DACs H1, D1) Van der Pauw circuit on diamond anvils was prepared for electrical measurements. Samples in all cells were heated with a pulsed infrared laser (1.06 μm) for 100-200 ms. Cerium particle prepared for loading into the DAC I1 was pre-compressed to 5-10 GPa in a dummy cell to produce a thin foil less than a 1μm thick. Thus, the hydride sample in the DAC I1 had a quasi-2D geometry. Samples H1 and D1 were prepared at Jilin University (JLU) without pre-compression procedure. DAC I1 was firstly investigated using the PPMS system in magnetic fields up to 16 T, then studied on the WM1 Water-Cooling Magnet (SHMFF) with a maximum magnetic field of 35 T, whose field sweep rate was 0.1-0.2 T/s. Finally, the same sample was investigated in pulsed magnetic fields at HLD HZDR up to 67 T. Samples H1 and D1 were investigated only in magnetic fields up to 28 T at SHMFF. The pressure in all DACs was not measured during the cooling process due to the lack of optical access. General information about prepared DACS can be found in Supporting Table S1.

Preliminary studies of the DAC I1 (Figure 1, 130 GPa), performed in magnetic fields up to 16 T, made it possible to determine the best combination of electrodes for voltage-current (*V-I*) measurements, to establish the presence of two superconducting crystaline phases in the sample with the critical temperatures ($T_C$) of 116-117 K (phase-I, CeH$_{10}$), and 100-102 K (phase-II, presumably, *hcp*-CeH$_9$), and the slope of the $B_{C2}(T)$ in the vicinity of the critical temperature ($dT_C/dB$) of –2.03 K/T for the phase-I (CeH$_{10}$) and –2.27 K/T for the phase-II (CeH$_9$). These values allow us to give an upper bound of the upper critical field $B_{C2}(0)$ using a linear extrapolation. The expected upper critical magnetic field for CeH$_{10}$ does not exceed 57 T, whereas for CeH$_9$ it is below 40 T. As we will see below, the superconductivity of CeH$_9$ can indeed be completely suppressed in much lower steady magnetic fields.

Use of the Bloch-Grüneisen [13-14] fit makes it possible to roughly estimate the Debye temperature in the DAC I1 sample as $T_\theta$ = 660 K, and the electron-phonon coupling strength $\lambda \approx 2.5$. The rather small value of the Debye temperature at this pressure, as well as nonlinear behavior of $R(T)$ in the interval of 150-200 K makes us recall another unusual hydride superconductor, SnH$_4$, which we studied earlier [9]. The quasi-linear dependence of electrical resistance on temperature in CeH$_{10}$ continues down to 0.18 $T_\theta$ (Supporting Figure S12).

X-ray diffraction study (SSRF, $\lambda$ = 0.6199 Å) from an asymmetric electrical DAC I1 of 70 mm long and 15 mm in diameter, prepared for pulsed magnetic fields, provides an extremely limited amount of information, namely gives us only two diffraction rings (Figure 1d), the first of which can be attributed to the (111) reflection from the $Fm\overline{3}m$-CeH$_{10}$. The small residual resistance of the heterogeneous CeH$_{9-10}$ sample appears in some experiments and depends on the electrode pair chosen for the measurements (Figure 1a, b).

Within the first step of our experiment at SHMFF we measured Raman spectrum of DAC I1 and, then, cooled the diamond anvil cell down to 0.3 K. A pronounced two-step structure of the superconducting transition was confirmed. Raman spectrum showed slight decrease in pressure to 125 GPa. During this first cooling cycle, the cooling rate increased uncontrollably to 20-30 K/min, as a result, the superconducting transition was shifted to lower temperatures by 10-25 K (Supporting Figure S2). This leads to an important observation that placing thermometer far from the sample and fast cooling/warming may easily led to unreliable and unreproducible values of the $T_C$.

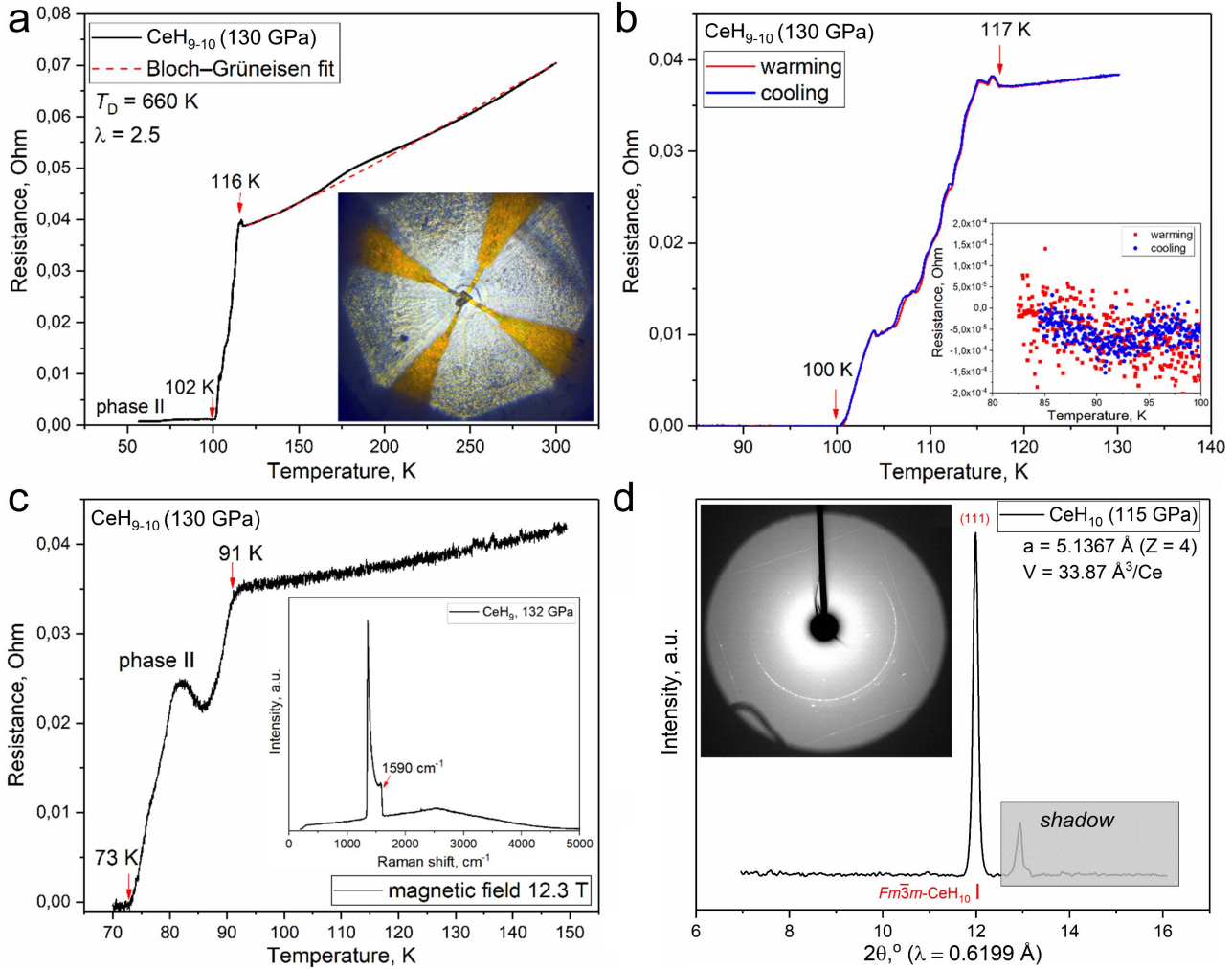

**Figure 1.** Dependence of the electrical resistance of $CeH_{9-10}$ (DAC I1) at 130 GPa on temperature and applied magnetic field, studied at the V.L. Ginzburg center for high-temperature superconductivity and quantum materials (LPI, Moscow). (a) Temperature dependence of the electrical resistance of a $CeH_{9-10}$ sample in zero magnetic field (delta mode, see Supporting Information). Inset: photograph of a pre-compressed Ce particle loaded with $NH_3BH_3$ prior to the laser heating. (b) Temperature dependence of the electrical resistance of the $CeH_{9-10}$ sample in the warming (red curve) and cooling (blue curve, rate is 2 K/min) cycles on another pair of contacts. Inset: residual sample resistance at $T < T_C$. (c) Pronounced two-phase pattern of the superconducting transition in a magnetic field of 12.3 T. Inset: Raman spectrum of the DAC I1, pressure is 130-132 GPa [15]. (d) X-ray diffraction pattern of the sample in DAC I1 measured several months after the experiment at 115 GPa. Pressure measurements in DAC I1 between experiments showed its gradual decrease over time. Due to the small opening angle, only one XRD peak can be observed. The peak corresponds to the unit cell parameter a = 5.137 Å of $Fm\bar{3}m$-$CeH_{10}$ at 115 GPa [7].

In the next series of the magnetic field sweeps from 0 to 33.2 T, we studied the behavior of the superconducting transitions in magnetic fields at temperatures from 0.4 to 120.6 K (Figure 2a). As a result, superconductivity was suppressed in the second phase ($P6_3/mmc$-$CeH_9$) and the obtained data made it possible to experimentally establish the upper critical field for $CeH_9$ as $B_{C2}(0) = 35$ T, which is higher than it was expected on the basis of Werthamer–Helfand–Hohenberg model [7,10-12]. Measured temperature dependence $B_{C2}(T)$ deviates from the linear fit $B_{C2}(T) = a \times (T_C - T)$ only at temperatures below 20 K (Figure 2b). The resulting $B_{C2}(0)$ value is between what the WHH model predicts (30 T) and what linear extrapolation gives (40 T). In general, the behavior of CeH$_9$ is in satisfactory agreement with the WHH model traditionally used for BCS superconductors.

Investigation of the voltage-current characteristics and the dependence of the critical current ($I_C$) on the applied magnetic field (Supporting Figures S6-S7) shows that the CeH$_9$ can be described in terms of the Bean-Kim model [16]: $I_C^{-1} \propto B$. The Hall effect measurements (see Supporting Figure S8) for DAC I1 lead to the Hall coefficient of 1.05-1.06×10$^{-4}$ Ω/T and the corresponding charge carriers (electrons) concentration of about 2–6×10$^{28}$ m$^{-3}$ in CeH$_{9-10}$. The error in determining the carrier concentration is due to the uncertainty of the sample thickness. In addition, the contribution of ionic (H$^-$) and hole conductivity (e.g., SnH$_4$ [9]) in CeH$_{9-10}$ cannot be discounted. Nevertheless, found concentration of electrons is similar to the carrier's concentration in common metals and corresponds to 1-3 electrons per unit cell of CeH$_{9-10}$. Now we can calculate the Fermi wave vector $k_F = (3\pi^2 n_e)^{1/3}$ and the Fermi wavelength $\lambda_F = 2\pi/k_F = 5 - 7.5$ Å, which is much lower than the sample thickness. Thus, the system should be considered as a fully three-dimensional.

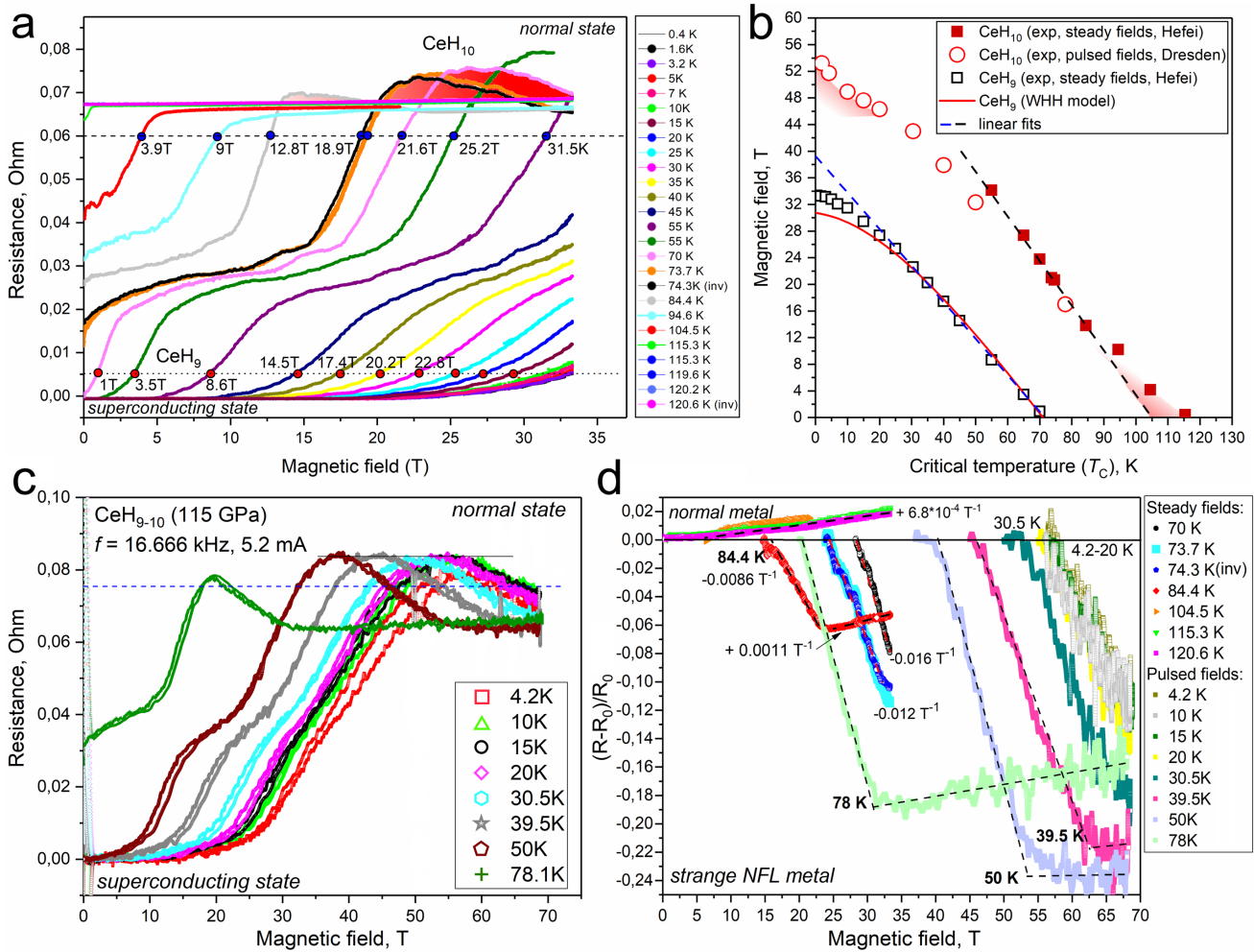

**Figure 2.** Dependence of the electrical resistance of CeH$_{9-10}$ (DAC I1) on the external magnetic field $R(B)$ measured in the AC mode at frequencies of 16.666 kHz (pulsed field) and 13.333 Hz (steady field). (a) Dependence of the electrical resistance of CeH$_{9-10}$ sample on the magnetic field at temperatures from 0.4 K to 120.6 K at 125 GPa measured at SHMFF. "Inv" in the legend means the reversal direction of the magnetic field. The anomalous area is indicated in red. (b) Joint magnetic phase diagrams of CeH$_9$ and CeH$_{10}$ and temperature dependence of the upper critical field $B_{C2}(T)$. $B_{C2}$ for CeH$_9$ were determined from the point of intersection of R(B) with the line $R(7.5\%) = 0.005$ Ω, and $B_{C2}$(CeH$_{10}$) was determined from the point of intersection of $R(B)$ with the line $R(90\%) = 0.06$ Ω. (c) The same experiment in pulsed magnetic fields up to 68 T (HLD HZDR) at 115 GPa. The visible hysteresis corresponds to sweep up and sweep down of the magnetic field during the pulse. The $R(90\%) = 0.06$ Ω line was used to extract $B_{C2}$(CeH$_{10}$). (d) Magnetoresistance $\rho = (R-R_0)/R_0$ versus applied magnetic field at different temperatures. $R_0$ corresponds to the normal state resistance right above the onset $T_C$ (see horizontal dash in panel (c)). For negative MR, $R_0$ is the resistance maximum.

Already in Figure 2a, one can notice a completely unexpected behavior of CeH$_{9-10}$ in the non-superconducting state when the superconductivity is suppressed by a strong magnetic field. Above 94.6 K, the $R(B)$ dependence in the normal state has no features, but at $T < 94.6$ K, the situation changes dramatically. At temperature of 84.4 K, the magnetoresistance (MR) of the sample is already negative and almost 20 times larger than in the normal state at temperature above the critical one (Figures 2d, 3a). At this temperature, even more interesting behavior is observed, namely: the sign of the magnetoresistance (MR = $d\rho/dB$) changes sharply from negative to positive at a magnetic field of 23–24 T. This behavior is completely abnormal for ordinary metals and may indicate transition to a pseudogap phase of the Ce superhydrides. At the same time, above 104 K CeH$_{9-10}$ exhibits behavior characteristic of many usual metals (Figures 2a, d). For example, in weak magnetic fields the MR depends quadratically on the field: $\rho = (R-R_0)/R_0 \propto \mu^2 B^2$, where $\mu$ is the charge carrier mobility (Supporting Figures S10 and S13, $\mu \approx 29$–43 cm$^2$/s·V). It should be noted that in strong fields, this dependence becomes linear $\rho \propto B$. We have previously observed a similar linear magnetoresistance for tin hydride SnH$_4$ [9] and lanthanum-neodymium hydrides (La,Nd)H$_{10}$ [17].

Simultaneously with the negative magnetoresistance, another anomaly is observed in the DACS I1: the sign of the temperature coefficient of electrical resistance in the normal (at $B > B_{C2}$) state changes: see the slope of red curve in Figure 3a. At temperatures above 94.6 K, the electrical resistance decreases with decreasing temperature (Figures 2a, 3a). This is a common behavior for metals in which the electrical resistance is caused by the scattering of electrons by phonons in the crystal lattice. But below 94.6 K, the situation changes: the resistance begins to increase sharply with decreasing temperature at $B > B_{C2}$, despite the negative magnetoresistance. This behavior is distinctive characteristic of semiconductors, strange non-Fermi liquid (NFL) metals, and the pseudogap phase of cuprate superconductors [18]. Similar demeanor has been observed previously for tin hydride SnH$_4$ [9], (La,Ce)H$_9$ [19] and in sulfur hydride H$_3$S when the pressure in a diamond anvil cell is reduced [5].

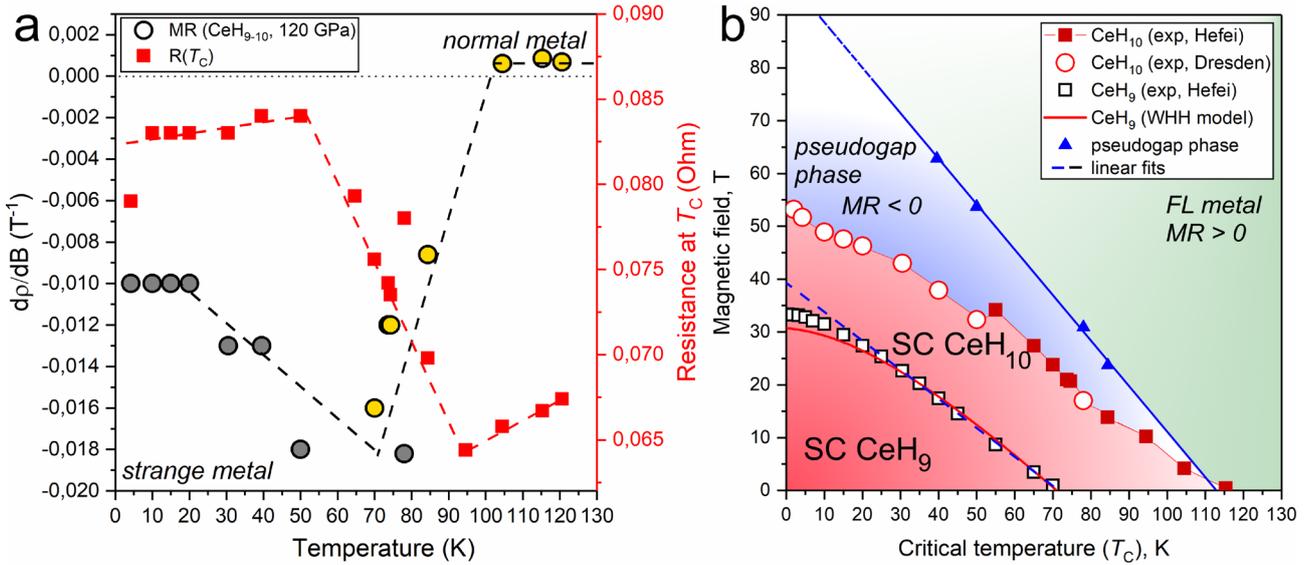

**Figure 3.** Coefficient of linear magnetoresistance ($d\rho/dB$) of the CeH$_{9-10}$ (DAC I1) at various temperatures and full magnetic phase diagram of cerium superhydrides. (a) Sign change of the magnetoresistance coefficient $d\rho/dB$ during the transition from the pseudogap to the normal Fermi liquid metal (FL) phase in a warming cycle from 2 K to 120.6 K. Yellow circles correspond to SHMFF experiments in steady fields, grey circles mark HLD HZDR experiments in pulsed magnetic fields. Red squares indicate normal state resistance right above the onset $T_C$. In many cases this R($T_C$) corresponds to maximum electrical resistance of the sample. (b) Full magnetic phase diagram of CeH$_9$ and CeH$_{10}$ at pressures of 115-125 GPa. "SC" (red) marks superconducting region, "pseudogap phase" (blue) corresponds to region with negative magnetoresistance and $dR/dT < 0$ at $B > B_{C2}$, "FL metal" denotes the Fermi liquid metal behavior of the sample with MR > 0 and positive $dR/dT > 0$ at $B > B_{C2}$.

Unusual behavior of the magnetoresistance was the reason for further studies of the sample in DAC I1 in stronger pulsed magnetic fields up to 68 T (Figure 2c, 115 GPa). Due to the movement of the DAC I1 over long distances, time gap, and many cooling and heating cycles, the pressure in the cells continued to decrease and amounted to 115 GPa in the HLD HZDR cycle of experiments. Pulse measurements allowed us to confirm the presence of a region of pronounced negative magnetoresistance at temperatures below 85–90 K and obtain additional inflection points where MR changes sign (Figure 2d). These points limit a separate region of the magnetic phase diagram (MR < 0, Figure 3b), which corresponds to the pseudogap phase of cerium superhydride $CeH_{10}$. We also were able to completely suppress superconductivity in the high-$T_C$ hydride $CeH_{10}$ (phase I) and complete the magnetic phase diagram for this compound (Figure 3b). The behavior of $B_{C2}(T)$ shows a pronounced deviation both from the WHH model and from the linear model $B_{C2}(T) \propto (T_C - T)$. It also has anomalies both in the region of strong (> 44 T) and weak (< 12 T) magnetic fields. Thus, cerium decahydride $CeH_{10}$ demonstrates pronounced deviations from the conventional behavior of metals and BCS superconductors both for magnetoresistance and electrical resistance $R(T)$, as well as for the upper critical magnetic field $B_{C2}(T)$.

*Low-field experiments with $CeH_9$ and $CeD_9$*

In order to separate the effects of hexagonal (*h*) $CeH_9$ from the behavior of cubic $CeH_{10}$, and to study the dependence of the upper magnetic field $B_{C2}(0)$ on the sample preparation procedure, we prepared high-pressure diamond anvil cells H1 (112-116 GPa) and D1 (128 GPa). The X-ray diffraction study at 112 GPa (DAC H1, Figure 4a) shows that the sample consists of two previously studied phases, $I4/mmm$-$CeH_4$ [6] and $P6_3/mmc$-$CeH_9$ [7], where the last phase dominates in the composition of the sample. The sample exhibits a superconducting transition at $T_C$ (onset) = 90 K (116 GPa, inset in Figure 4d), however, the resistance does not drop to zero due to the presence of the parasitic $CeH_4$ phase. A similar problem exists in the DAC D1, where a 2-fold drop in electrical resistance is observed at $T_C(CeD_9)$ = 67 K at 111 GPa (Supporting Figure S15). The sample in DAC D1 is also a mixture of $CeD_9$ and $CeD_4$ (about 1:1, Supporting Figure S16). The drop of electrical resistance in DAC D1 corresponds to appearance of superconductivity in $CeD_9$ and makes it possible to estimate the isotope coefficient $\alpha = \ln[T_C(CeH_9)/T_C(CeD_9)]/\ln 2 = 0.43$ at 111-116 GPa. This value is quite close to that predicted in the framework of the BCS theory (0.48-0.5) and agrees with previous results [7].

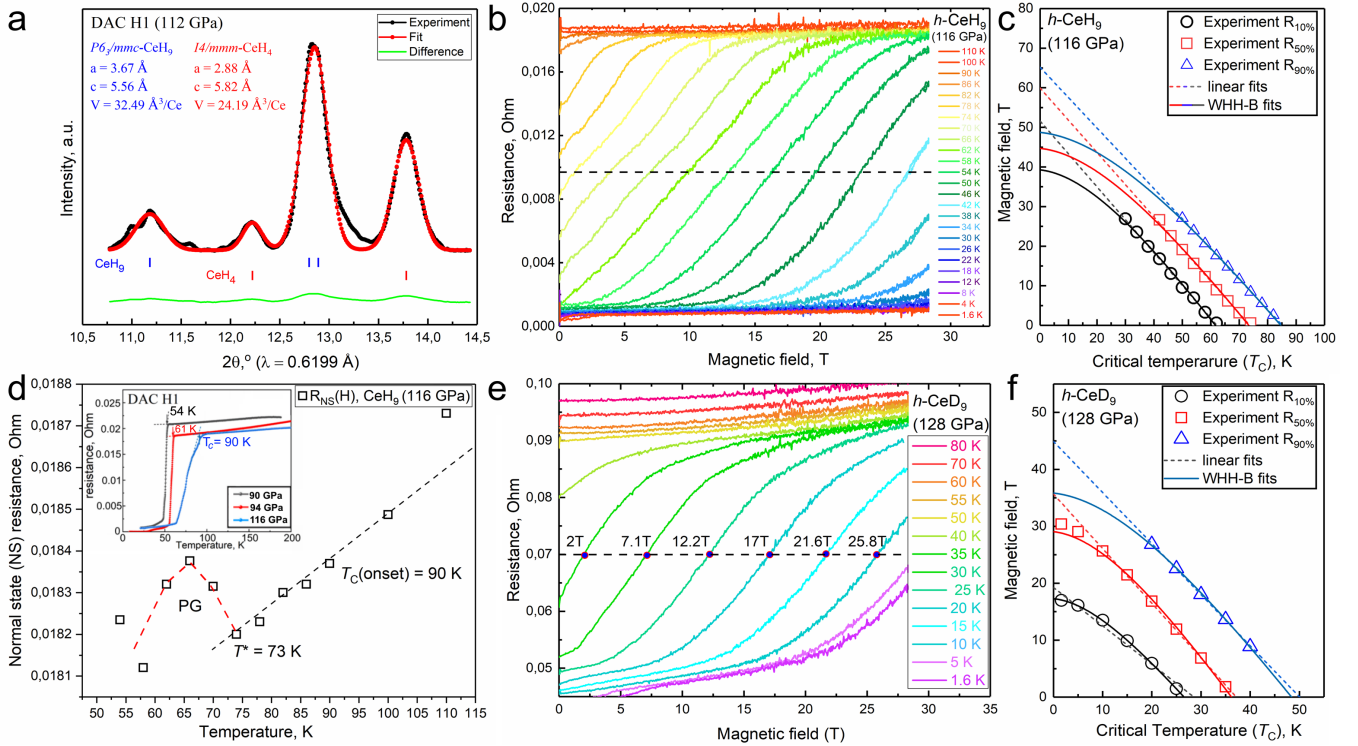

**Figure 4.** Results of studying samples of $CeH_9$ and $CeD_9$ in DACs H1 and D1 at 112-128 GPa. (a) Experimental XRD pattern and the Le Bail [20] refinement of the unit cell parameters of $I4/mmm$-$CeH_4$ and $P6_3/mmc$-$CeH_9$ at 112 GPa. Measurements were done at SSRF (λ = 0.6199 Å). Despite the presence of two phases, the superconducting transition in the DAC H1 is due to $CeH_9$ phase. (b) Dependences of electrical resistance on steady magnetic field of the $CeH_9$ sample (DAC H1) for temperatures from 1.6 to 110 K (SHMFF). (c) Magnetic phase diagram of $CeH_9$ at 116 GPa and extrapolation of the upper critical field to low temperatures (WHH-B means the WHH model simplified by Baumgartner[21]). (d) The jump of the $CeH_9$ electrical resistance in the normal state R($T_C$) in a magnetic field at $T^* < T_C$, which may correspond to the boundary between the superconducting (SC) and pseudogap (PG) phases. See also Supporting Figure S13. Inset: superconducting transitions in DAC H1 at pressures from 90 to 116 GPa. (e) Dependences

of the electrical resistance of sample in DAC D1 (CeD$_9$) on steady magnetic fields for temperatures from 1.6 to 80 K (SHMFF). (f) Full experimental magnetic phase diagram for CeD$_9$ at 128 GPa.

A limited study of samples H1 and D1 in magnetic fields up to 28 T shows that the properties of polyhydrides vary from sample to sample, may depend on the purity and its thickness. Thus, the upper critical field $B_{C2}(0)$ of CeH$_9$ in DAC H1 can reach 45-50 T (Figure 4c), which is 30% higher than in the DAC I1. This corresponds to higher defects concentration in heterogeneous CeH$_9$ + CeH$_4$ sample (D1) leading to stronger pinning of magnetic flux vortices and their greater resistance to an external magnetic field [22]. The low reproducibility of the upper critical field in hydride superconductors has been noted before (e.g., YH$_6$ [2-3]). Due to the limited available magnetic fields for BeCu DACs H1 and D1 (external cell's diameter was d = 23 mm), in the case of CeH$_9$ (in DAC H1) we could not go beyond the area of linear dependence of $B_{C2}(T) \propto (T_C - T)$ (Figures 4b, c), but for CeD$_9$ the superconductivity was suppressed at 30-31 T (Figure 4e) and the dependence $B_{C2}(T)$, in general, obeys the WHH model (Figure 4f).

As we see, CeH$_9$ and CeD$_9$ are close to what the WHH model predicts for $B_{C2}(T)$, while the CeH$_{10}$ sample in the DAC I1 obviously does not follow this model. Currently, there are several polyhydrides that do not obey the WHH model: this is SnH$_4$ [9], YH$_4$ [8], (Ca, Nd, Zr)H$_x$ [17], and cubic CeH$_{10}$ – is another representative of this group of compressed hydrides.

**Discussion**

Since the discovery of superconducting sulfur hydride H$_3$S in 2015 [5], polyhydrides have been considered to be a kind of benchmark of conventional electron-phonon mediated superconductivity. They exhibit the properties of metallic hydrogen [23], demonstrate a pronounced isotope effect [24], and lend themselves well to theoretical analysis [25-26]. However, already in the first work [5], some features of the temperature dependence of the electrical resistance were observed, which suggested the non-Fermi-liquid behavior of several polyhydrides [9]. In particular, the most obvious problem in describing the behavior of some polyhydrides in the normal state is the negative temperature coefficient of resistance $dR/dT < 0$. This situation has been observed for H$_2$S and H$_3$S [5], PH$_3$ [27], and (La, Ce)H$_9$ [19]. For ordinary metals, a decrease in the population of phonon states with decreasing temperature leads to a decrease in the scattering of electrons by phonons. It is part of the electron-phonon interaction and described by the Eliashberg transport function $\alpha^2_{tr}F(\omega)$. In most cases, this function is very close to the Eliashberg function $\alpha^2F(\omega)$, which describes superconductivity[28]. Therefore, the negative and zero temperature coefficient $dR/dT \propto \lambda$, where $\lambda$ is the electron-phonon coupling strength, observed in H$_3$S, PH$_3$ and (La,Ce)H$_9$, is incompatible with the concept of polyhydrides as ordinary metals, described within the Fermi liquid model [28]. Considering the quasi-linearity of the electrical resistance discussed above, we should say that the electron-phonon scattering does not play a main role in CeH$_{10}$.

Similar anomalies are also observed in the Hall effect and magnetoresistance of polyhydrides in pulsed magnetic fields [17]. For example, the change in the sign of the magnetoresistance and the Hall effect was recently found in Nd-doped LaH$_{10}$ [17]. It is interesting to note that the appearance of negative magnetoresistance and a jump in electrical resistance in strong magnetic fields below $T_C$ is observed in many cuprate superconductors such as electron-doped Nd$_{2-x}$Ce$_x$CuO$_4$ [29], Bi$_2$Sr$_2$CaCu$_2$O$_{8+y}$ [30-31], and Bi$_{1.6}$Pb$_{0.4}$Sr$_2$CaCu$_{1.96}$Fe$_{0.04}$O$_{8+\delta}$ [32]. The key idea that explains this behavior of superconductors in the normal state is the local fluctuations of the order parameter above $T_C$. Such fluctuations in a pseudogap phase are no longer able to provide a continuous supercurrent and the sample has a finite resistivity. But at the same time, they reduce the number of occupied states at the Fermi level and the number of conduction electrons due to local formation of a superconducting gap and, thus, increase the electrical resistance of samples [33].

Usually, the presence of a pseudogap is proved by measurements related directly to the energy spectrum: ARPES [34], tunnel measurements [35], heat capacity [36], nuclear magnetic resonance [37]. However, in the case of superhydrides stabilized by high pressure, it is immensely challenging to use these methods due to limited access to the sample and its microscopic size. For this reason, in this work, we focus on indirect signs of the pseudogap: anomalies in the behavior of the resistance $R(T)$, the appearance of a negative MR region, and the reversal of the sign of $dR/dT$.

In the absence of direct evidence of the pseudogap phase, let's consider what other physical phenomena can lead to similar anomalous transport properties of the $CeH_{9-10}$ samples. We first need to know the characteristic electron scattering time ($\tau$) in $CeH_{9-10}$, which can be estimated based on the Hall effect and magnetoresistance data (Supporting Figures S10, 13) in the normal (non-superconducting) state:

$$\tau = \frac{m_e \mu}{e} = \frac{m_e}{e^2 n_e \rho_s}, \quad (1)$$

where $e$ – is the electron charge, $m_e$ – is the electron mass, $\mu$ – is mobility of electrons (30-40 cm$^2$/V·s in our case), $\rho_s$ – the sample resistivity (2–5×10$^{-7}$ Ohm*m, see Supporting Information), $n_e$ – is the charge carrier concentration (2–6×10$^{28}$ m$^{-3}$, taken from the Hall effect). This leads to $\tau \approx 3 \cdot 10^{-15}$ s or $\hbar/\tau = 0.22$ eV in agreement with the values obtained for other superhydrides [9,17]. This corresponds to the properties of "dirty" metals. Taking into account that the Fermi energy $E_F$ for hydrides is about 2-10 eV and higher, we come to the conclusion that

$$\frac{E_F \tau}{\hbar} \gg 1, \quad \frac{k_B T_C}{\hbar/\tau} \ll 1. \quad (2)$$

This excludes the effects associated with the metal-insulator transition and hopping transport from consideration, since the system has pronounced metallic properties. The contribution of the electron-electron scattering with Umklapp processes involved (usual Fermi-liquid effects) should not play a significant role, since it depends on temperature as $\propto T^2$, while the effects under consideration have a comparable amplitude $\Delta R/R$ both at 4 K and at 50 K. Contribution of the electron-electron interaction corrections in impurity scattering (Altshuler-Aronov effect [38]) should result in $\Delta R/R \propto \sqrt{T}$ dependence [39] which is inconsistent with our experimental data.

Finally, we consider magnetic effects that may lead to alternative explanations for the negative MR in $CeH_{9-10}$. First of all, according to our knowledge, all spin-polarized theoretical calculations unambiguously testify in favor of the absence of macroscopic magnetic ordering in cerium polyhydrides [40-43]. Therefore, scattering by fluctuations of the magnetic moment in the vicinity of the Néel or Curie points must be excluded. However, a significant (by a factor of 2) decrease in the maximum $T_C$ of $CeH_9$ and $CeH_{10}$, compared to isostructural $LaH_9$ and $LaH_{10}$, which is due to anisotropic scattering of Cooper pairs on Ce atoms, can be considered as manifestation of the microscopic magnetic moments of cerium atoms [17]. Theoretically, in the vicinity of the spin density wave (SDW), such fluctuations can also take place, but all these effects should be better manifested in more inhomogeneous $CeH_9$ + $CeH_4$ samples (DACs H1 and D1) with higher $B_{C2}(0)$. In fact, the opposite situation is observed. The anomalies are most pronounced in the best sample of DAC I1. Furthermore, considering the non-linear effects observed in the anomalous Hall effect measurements (Supporting Figure S8), one can estimate the expected saturation field of the Ce magnetic moments (and potential impurities) as 10-12 T. However, the negative MR in $CeH_{9-10}$ is also observed at 40-60 Tesla (Figure 2c), much higher than the saturation field of magnetic impurities. This fact excludes explanations associated with ferromagnetic and antiferromagnetic magnons. Finally, it is noteworthy to mention that the negative MR in our experiments is tied to the superconducting transition, and this correlation is clearly nonrandom. All effects associated with the magnetism of Ce atoms should be tied to the magnitude of the magnetic field, which is not the case.

Pseudogap phase is not an attribute of only cuprate superconductors. For BCS superconductors it also occurs, but due to the low critical temperature, the superconducting fluctuations are weakly pronounced. For instance, the pseudogap phase was found in TiN thin films [35] and disordered NbN samples [44-45]. Thus, building a bridge between high-temperature superconducting cuprates and low-temperature BCS superconductors, we come to conclusion that cerium superhydrides have a complex phase diagram, one of the parts of which is a region of superconducting fluctuations above $T_C$, similar to the pseudogap phase of cuprates.

**Conclusions**

We studied parameters of the superconducting state, voltage-current characteristics, magnetoresistance and the Hall effect in $CeH_{10}$, $CeH_9$ and $CeD_9$ under pressure of 112-130 GPa in strong steady and pulsed magnetic fields up to 68 T. Experimental magnetic phase diagrams down to 0.4 K were constructed for all compounds and their upper critical fields were established. The experimental behavior of $B_{C2}(T)$ of hexagonal CeH$_9$ and CeD$_9$, in general, can be described in terms of the WHH model. Cubic CeH$_{10}$ does not obey the WHH model, it has a bilinear $B_{C2}(T)$ with anomalies at low temperatures and in the vicinity of $T_C$.

Moreover, cerium superhydride CeH$_{10}$ exhibit pronounced properties of a strange metal below $T_C$: large $B$-linear negative magnetoresistance, the reversal of the sign of temperature coefficient of electrical resistance ($dR/dT$) below 85-95 K, and quasi-linear $R(T)$ in non-superconducting state, which corresponds to the properties of the pseudogap phase of cuprate superconductors. Signs of such behavior are also found in CeH$_9$ below 73 K. This experiment shows that the physics of high-$T_C$ superhydrides is far from trivial, and the properties of these compounds are rather close to those of cuprates.

**Acknowledgements**

This work was supported by the National Key R&D Program of China (Grant No. 2022YFA1405500), the National Natural Science Foundation of China (Grants No. 52072188, and No. 11974133), the Program for Changjiang Scholars and Innovative Research Team in University (Grant No. IRT_15R23) and Zhejiang Provincial Science and technology innovation Team (2021R01004). A portion of this work was performed on the Steady High Magnetic Field Facilities (SHMFF), High Magnetic Field Laboratory, Chinese Academy of Science (CAS). This work was supported by HLD HZDR, member of the European Magnetic Field Laboratory (EMFL). The high-pressure experiments were supported by the Ministry of Science and Higher Education of the Russian Federation within the state assignment of the FSRC Crystallography and Photonics of the RAS and by the Russian Science Foundation (Project No. 22-12-00163). The research used resources of the LPI Shared Facility Center. V.M.P, A.V.S. and O.A.S. acknowledge the support of the state assignment of the Ministry of Science and Higher Education of the Russian Federation (Project No. 0023-2019-0005). The authors thank the staff of the Shanghai Synchrotron Radiation Facility for their help during the synchrotron XRD measurements. The authors express their gratitude to Prof. Viktor Struzhkin (HPSTAR) and Dr. Thomas Meier (HPSTAR) for technical and financial support.

**Contributions**

D.S., J.G., D.Z., W.C., T. H., A. S., O. S., C. X. and I. T. performed the experiments. D. S. and A.K. prepared the theoretical calculations. D.S., V.P., V.S., and X.H. analyzed the data and wrote the paper. D.S. and X.H. conceived this project. All the authors discussed the results and offered useful inputs.

**Data availability**

The authors declare that the main data supporting our findings of this study are contained within the paper and Supporting Information. All relevant data are available from the corresponding authors upon request. The raw data can be found on the github: https://github.com/mark6871/CeH10-pseudogap-paper/tree/main

**Code availability**

Quantum ESPRESSO and EPW codes are free for academic use and available after registration at https://www.quantum-espresso.org/ and https://epw-code.org/, respectively.

# SUPPORTING INFORMATION

## Evidence for Pseudogap Phase in Cerium Superhydrides: CeH$_{10}$ and CeH$_9$

Dmitrii Semenok[1], Jianning Guo[2], Di Zhou[1, *], Wuhao Chen[2], Toni Helm[3], Alexander Kvashnin[4], Andrei Sadakov[5], Oleg Sobolevsky[5], Vladimir Pudalov[5,6], Chuanyin Xi[7], Xiaoli Huang[2, *], and Ivan Troyan[8, *]

[1] Center for High Pressure Science & Technology Advanced Research, Bldg. #8E, ZPark, 10 Xibeiwang East Rd, Haidian District, Beijing, 100193, China

[2] State Key Laboratory of Superhard Materials, College of Physics, Jilin University, Changchun 130012, China

[3] Hochfeld-Magnetlabor Dresden (HLD-EMFL) and Würzburg-Dresden Cluster of Excellence, Helmholtz-Zentrum Dresden-Rossendorf (HZDR), Dresden 01328, Germany

[4] Skolkovo Institute of Science and Technology, Skolkovo Innovation Center, Bolshoy Boulevard, 30/1, Moscow 121205, Russia

[5] V. L. Ginzburg Center for High-Temperature Superconductivity and Quantum Materials, P. N. Lebedev Physical Institute, Russian Academy of Sciences, Moscow 119991, Russia

[6] National Research University Higher School of Economics, Moscow 101000, Russia

[7] Anhui Province Key Laboratory of Condensed Matter Physics at Extreme Conditions, High Magnetic Field Laboratory of the Chinese Academy of Science, Hefei 230031, Anhui, China.

[8] Shubnikov Institute of Crystallography, Federal Scientific Research Center Crystallography and Photonics, Russian Academy of Sciences, 59 Leninsky Prospekt, Moscow 119333, Russia

Corresponding authors: Di Zhou (di.zhou@hpstar.ac.cn), Xiaoli Huang (huangxiaoli@jlu.edu.cn) and Ivan Troyan (itrojan@mail.ru)

## Content

# 1. Methods

*Experiment*

The crystal structure was determined using the synchrotron X-ray diffraction (XRD) on the BL15U1 synchrotron beamline with a wavelength of 0.6199 Å at the Shanghai Synchrotron Research Facility (SSRF). The experimental XRD images were integrated and analyzed for possible phases using the Dioptas software package [1]. To fit the diffraction patterns and obtain the cell parameter, we analyzed the data using Materials Studio and Jana2006 software [2], employing the Le Bail method [3].

Magnetoresistance measurements under high pulsed magnetic fields were carried out in a 24 mm bore of 72 T resistive pulse magnet (rise time of 15 ms) at the Helmholtz-Zentrum Dresden-Rossendorf (HLD HZDR). Strands of the Litz wire glued to the silver paint were moved closer together to minimize open loop pickup. All twisted pairs were fixed using the GE7031 varnish. A bath helium cryostat was used, which made it possible excellent control of the DAC I1 temperature between 1.6 and 60 K. A 30 cm long NiCr wire with a resistance of about 100 Ohms wrapped around the diamond chamber was used as a heater. Cernox thermometers were attached to the DAC I1 body for measurements of the temperature. A high-frequency (3.333 and 16.666 kHz) lock-in amplifier technique was employed to measure the sample resistance. For the measurements in high magnetic fields, we used a four-probe AC method with the excitation current of 5-10 mA, the voltage drop across the sample was amplified by an instrumentation amplifier and detected by a lock-in amplifier. Comparing the up and down sweep resistance curves at various field sweep rates, no significant sample heating was observed during ~ 150 ms long magnet pulse at temperatures above 35 K (see Figure S1). In general, we used the same methodology as in the previous studies of (La, Nd)H$_{10}$ [4] and SnH$_4$ [5].

Low temperature resistivity AC measurements were performed between 0.4 K and 73 K using 3He system, in applied fields up to 35 T, the sweep rate was 0.1 T/s. The temperature in the system was measured using a Lakeshore 336 temperature controller and two Cernox 1030 thermometers (error is ±0.2 K). Four Au/Ta electrodes were sputtered on the diamond anvil of DAC I1. For DACs H1 and D1 we used Mo electrodes with thickness of about 200-300 nm (Table S1). In the experiment with steady magnetic fields at SHMFF, we used sinusoidal wave with frequency of 13.333 Hz and the current amplitude of 2 mA (Keithley 6221 current source). Standard Research 830 lock-in amplifier was used to amplify the AC signal. Cernox thermometers were mounted on a variable temperature inset (VTI) in 15 mm above the sample. Due to the limited size of the magnet bore (17 mm), it was not possible to place thermometers directly on the diamond anvil cell. Sample in the DAC was placed in 9.2 cm from the bottom of cryostat and in 4.5 cm from the magnetic field center. As a result, the effective magnetic field in the sample was 5% lower than the nominal value (maximum is 35 T). The high field transport measurements were performed using the WM1 Water-Cooling Magnet at the High Magnetic Field Laboratory of the Chinese Academy of Sciences (SHMFF, Hefei, China, see Figure 5).

**Table S1**. List of high-pressure DACs used in this experiment, samples loaded into them and products obtained.

| DACs and their diameter | Experiment | Sample | Products |
|---|---|---|---|
| DAC I1 (d = 15.3 mm) | SHMFF 2023 (Hefei, China) HLD HZDR 2023 (Dresden, Germany) | Ce/NH$_3$BH$_3$ | $Fm\bar{3}m$-CeH$_{10}$ + $P6_3/mmc$-CeH$_9$ |
| DAC H1 (d = 23 mm) | SHMFF 2023 (Hefei, China) | Ce/NH$_3$BH$_3$ | $P6_3/mmc$-CeH$_9$ + $I4/mmm$-CeH$_4$ |
| DAC D1 (d = 23 mm) | SHMFF 2023 (Hefei, China) | Ce/ND$_3$BD$_3$ | $P6_3/mmc$-CeD$_9$ + $I4/mmm$-CeD$_4$ |

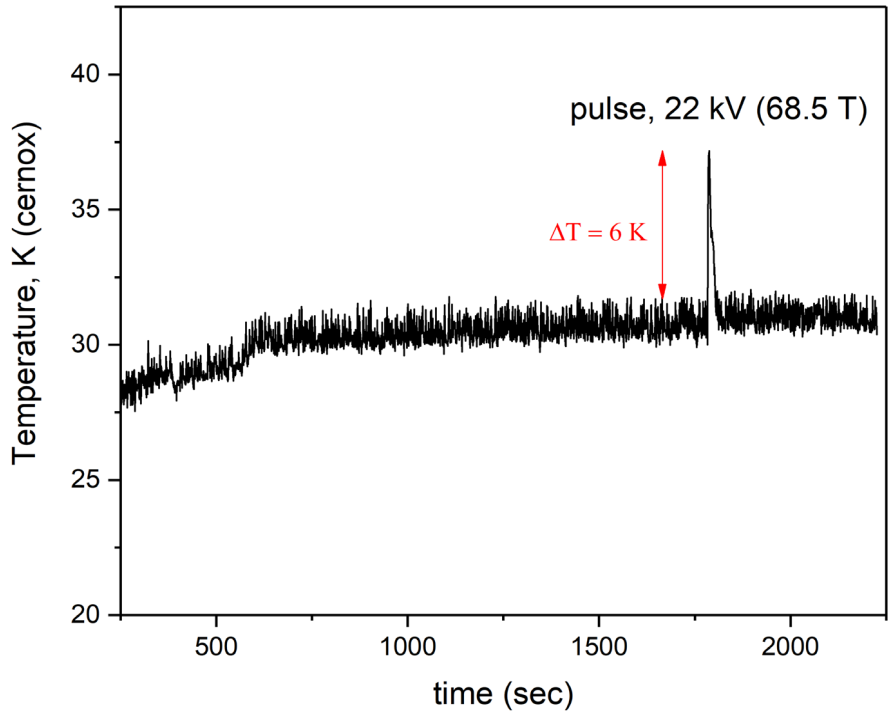

**Figure S1.** Time dependence of the sample (DAC I1) temperature during a pulsed experiment with a magnetic field of 68 T (pulse of 22 kV, HLD HZDR) at 30 K. It can be seen that due to electromagnetic induction, a sharp "phantom" temperature jump of 6 K initially occurs. However, heating by eddy currents corresponds to the wider part of the peak of approximately 3 K height. In any case, due to the limited rate of heat transfer between the metal body and the diamond anvils, the sample temperature does not seriously change during the pulse (150 ms).

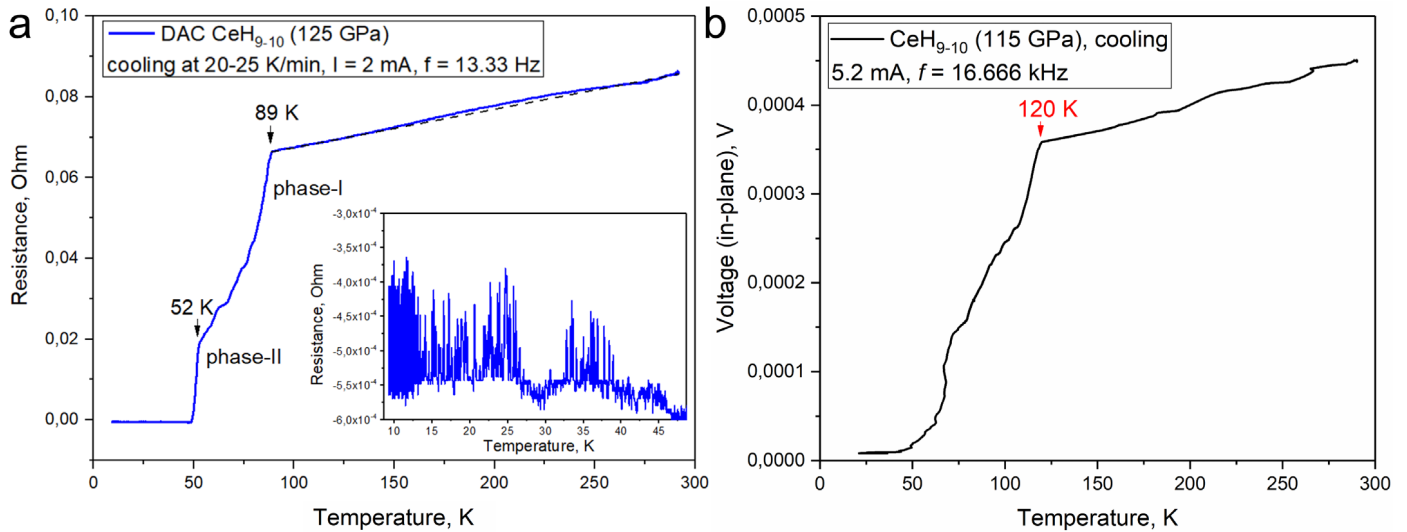

**Figure S2.** Temperature dependence of the electrical resistance of the $CeH_{9-10}$ sample at a very high sample cooling rate of 20-25 K/min in AC mode. (a) experiment at the SHMFF, (b) experiment at the HLD HZDR. Within the HLD HZDR experiment the cernox thermometer was placed on a seat of DAC, therefore, despite the high cooling rate, the detected critical temperature almost corresponds to the value (117 K) obtained with slow cooling (1-2 K/min)

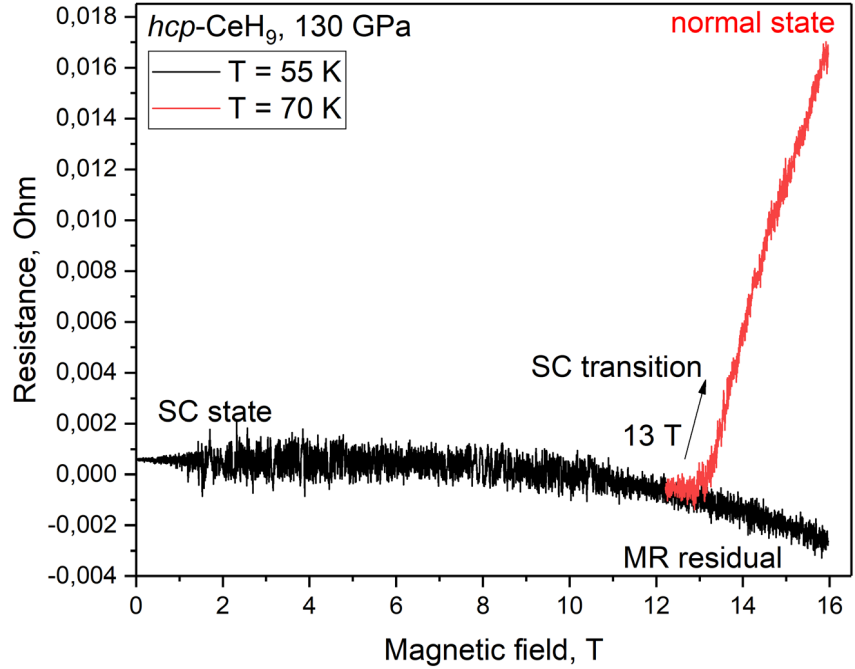

**Figure S3.** Dependence of the electrical resistance of the sample on the magnetic field at 70 K (red curve) and 55 K (black curve), DAC I1. At 70 K the superconducting state is destroyed at 13 T firstly in in CeH$_9$, while at 55 K the superconducting state cannot be destroyed. The slope of the curve at 55 K is due to the residual resistance of the sample on the selected pair of contacts (see Figure 1a). Experiment was done at LPI (Moscow).

The superconducting transition in CeH$_{9-10}$ (DAC I1) was also investigated for different measuring current frequencies in the fast heating and cooling mode using Physical Property Measurement System (PPMS, Quantum Design) system at the HLD HZDR (Figure 4). This experiment shows that the detection of a superconducting transition in DACs is also possible using a current with a frequency of 510 kHz and higher, which is an extremely important property for studies in superstrong (destructive) pulsed magnetic fields with a pulse duration of a few milliseconds or less [6].

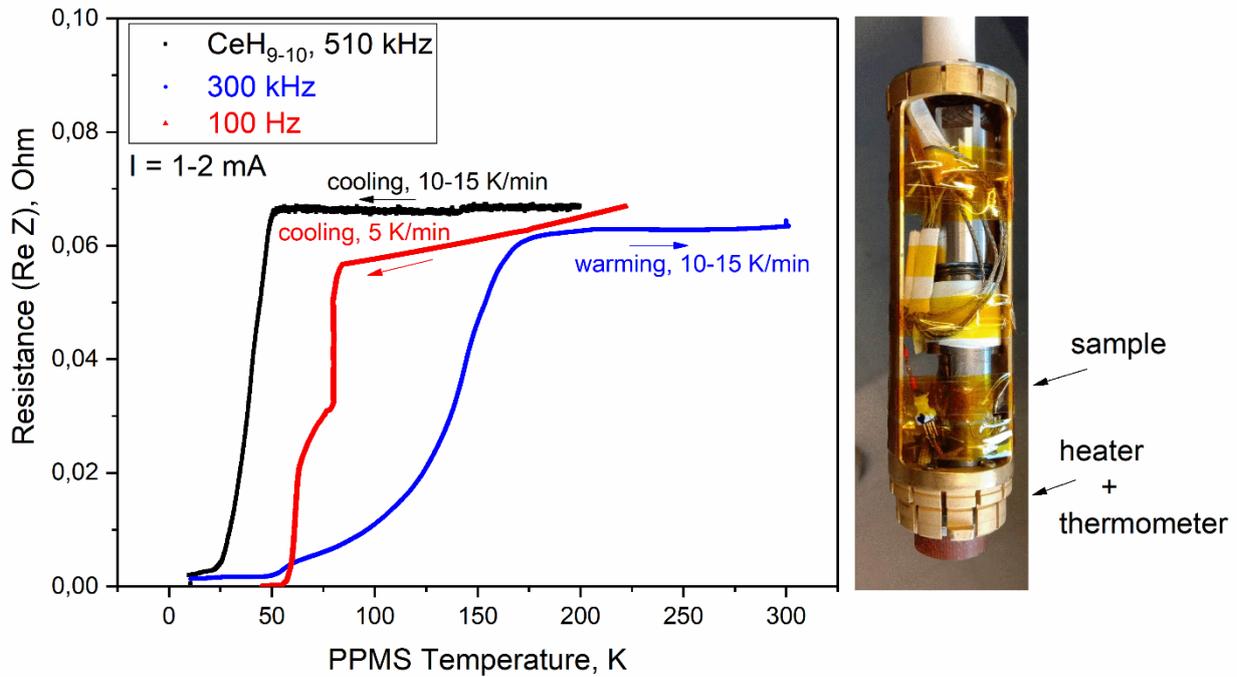

**Figure S4.** PPMS investigation of the transition to the superconducting state of the CeH$_{9-10}$ sample (DAC I1) using a measuring current of various frequencies (0.1 – 510 kHz, HZDR HLD). Due to the large contribution of frequency-dependent reactance (Imaginary impedance, Im Z), the $R(T)$ = Re $Z(T)$ dependences are scaled for convenience. The dependence of the observed transition temperature on the rate of cooling or heating is clearly visible due to the large distance between the sample and the thermometer in the probe we used.

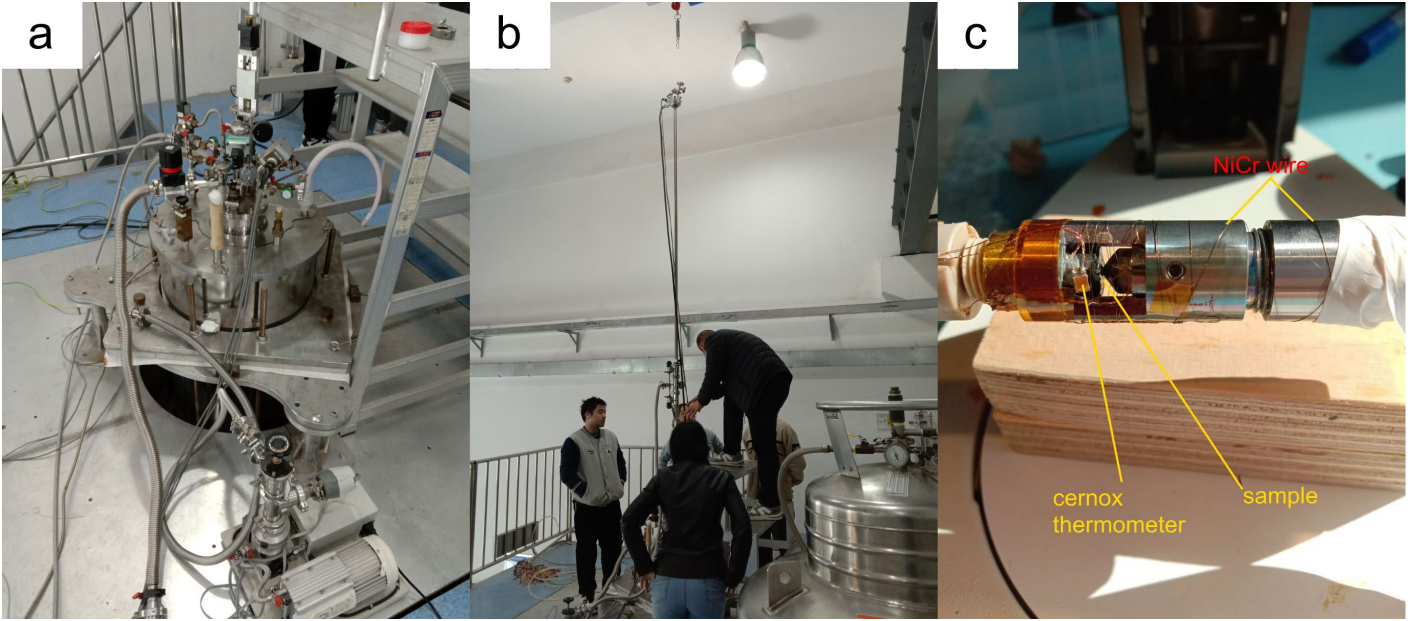

**Figure S5.** Water-cooled magnet WM1 at the SHMFF: (a) upper part including various ports, (b) the process of placing the DAC I1 into the magnet using a VTI of about 4.5 meters long. (c) View of the DAC I1 with the NiCr heater wire wound around the DACs body, and the temperature sensor glued on the seat near the $CeH_9$-$CeH_{10}$ sample. In this configuration the DAC I1 was used at the HLD HZDR.

*Theory*

In some cases, it is possible to estimate the Debye temperature $\theta_D$ of hydride superconductors from the dependence of the electrical resistance $R(T)$ on the temperature, using fit of the experimental $R(T)$ by the Bloch–Grüneisen (BG) formula [7-8]:

$$R(T) = R_0 + A\left(\frac{T}{\theta_D}\right)^5 \int_0^{\frac{\theta_D}{T}} \frac{x^5}{(e^x-1)(1-e^{-x})} dx, \qquad (S1)$$

where $A$, $\theta_D$, and $R_0$ were found using the least squares method. Recently, it has been demonstrated that when $R \rightarrow 0$ at $T < T_C$, the formula gives reasonable values of the Debye temperatures for $H_3S$ and $LaH_{10}$ [9]. The analysis of the experimental data for $P6_3/mmc$-$CeH_9$ (Figure 1 and Figure S19d) shows that $\theta_D$ and $\omega_{log}$ remain around 500–600 K at 130–160 GPa, which indicates the need of high $\lambda = 2 \pm 0.5$ to ensure that the observed critical temperature of superconductivity is above 100 K.

Calculations of superconducting and transport properties of $CeH_9$ were carried out using QUANTUM ESPRESSO (QE) package [10-11]. The phonon frequencies and electron–phonon coupling (EPC) coefficients were computed using the density functional perturbation theory [12], employing the plane-wave pseudopotential method with a cutoff energy of 70 Ry, and the Perdew–Burke–Ernzerhof exchange–correlation functionals[13]. Within the optimized tetrahedron [14] and the interpolation [15] methods, we calculated the electron–phonon coupling coefficients $\lambda$ and the Eliashberg functions.

## 2. Critical currents in *P6₃/mmc*-CeH₉

Due to the fact that measurements of critical currents require the boundary between the superconducting and non-superconducting states on the *V-I* characteristics to be as clear as possible, such measurements are possible only for the lowest-temperature superconducting transition, i.e., for *P6₃/mmc*-CeH₉ in DAC I1. If there is significant residual sample resistance (as in DACs H1 and D1) these measurements are irrelevant.

Critical currents were studied at the SHMFF in the current sweep mode of a sinusoidal wave with a frequency of 13.33 Hz, the current amplitude increased from 0 to 30 mA (up to 10 mA at 69.6 K) and decreased back to zero (Figure S6). AC critical current measurements are noticeably worse than the delta mode measurements in terms of data quality. Delta mode – is a differential method for measuring electrical resistance with a change in the direction of flowing current, aimed at eliminating the influence of contact and thermoelectric phenomena on the resistance of the sample. Due to the presence of parasitic capacitance and inductance there is always a non-zero residual reactance and the phase shift in the superconducting state, which in our case was excluded by subtracting the *V-I* curve obtained in the zero field $B = 0$ at the corresponding temperature. Thus, the presented data are obtained as $V(I, B) = V_{exp}(I, B) - V_{exp}(I, 0)$.

The electrical current step was chosen to be 0.1–0.2 mA, each scan took about 2 minutes. The measurements were carried out in a constant magnetic field from 0 to 8 T. In most cases, in zero field, a current of 30 mA was not able to suppress superconductivity in CeH₉, except for the point of 69.6 K, where a linear increase in resistance is observed above 3-3.5 mA (Figure S6d). In other cases, to suppress superconductivity, the voltage-current characteristics were taken in a magnetic field. In many cases, a noticeable heating of the sample by the current was observed, which was expressed in the parabolic character of the *V(I)* dependencies (Figure S6c) and noticeable *V-I* hysteresis between current sweeps up and down.

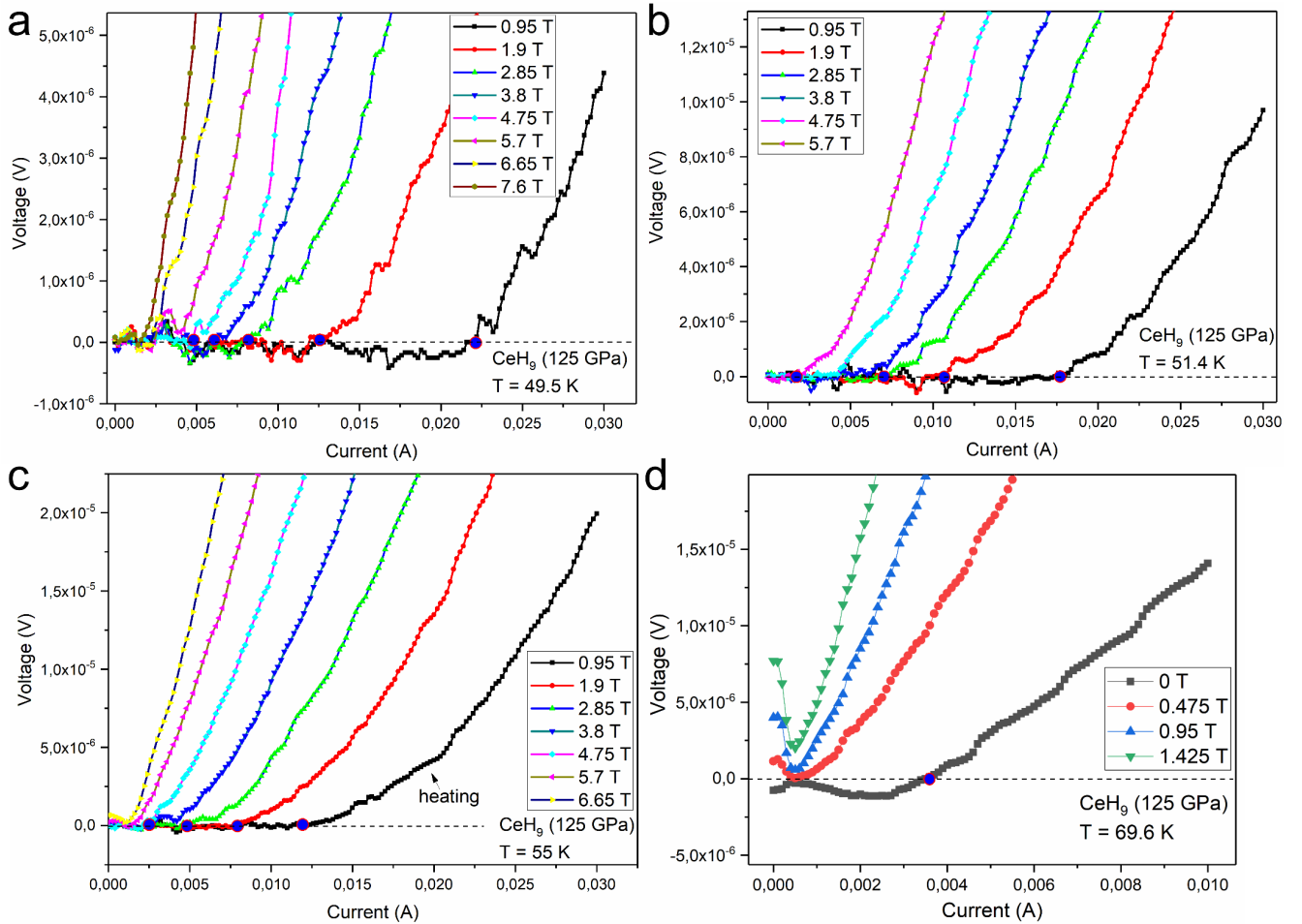

**Figure S6.** Current-voltage (*V-I*) characteristics and critical currents in hexagonal CeH₉ (DAC I1) in magnetic fields from 0 to 7.6 T at temperatures: (a) 49.5 K; (b) 51.4 K; (c) 55 K and (d) 69.6 K. In almost all cases, the suppression of superconductivity by electric current requires an external magnetic field. Due to the presence of reactance in the circuit, the resistance in the superconducting state is formally negative. To compensate for the reactance, we used the *V-I* characteristic in zero field as the background.

Another approach to the $I_C$ measurements was chosen in the experiment at the HLD HZDR facility. In that case we could not fix the magnetic field and it was decided to fix the current and detect *V-B* characteristics at three temperature points: 40 K, 45 K and 50 K, using weak magnetic field pulses up to 10 Tesla (Figure S7). The residual resistance (the real impedance, Re Z) in the superconducting state was corrected to zero by the phase multiplier ($e^{i\varphi}$, $\varphi$ - within a few degrees).

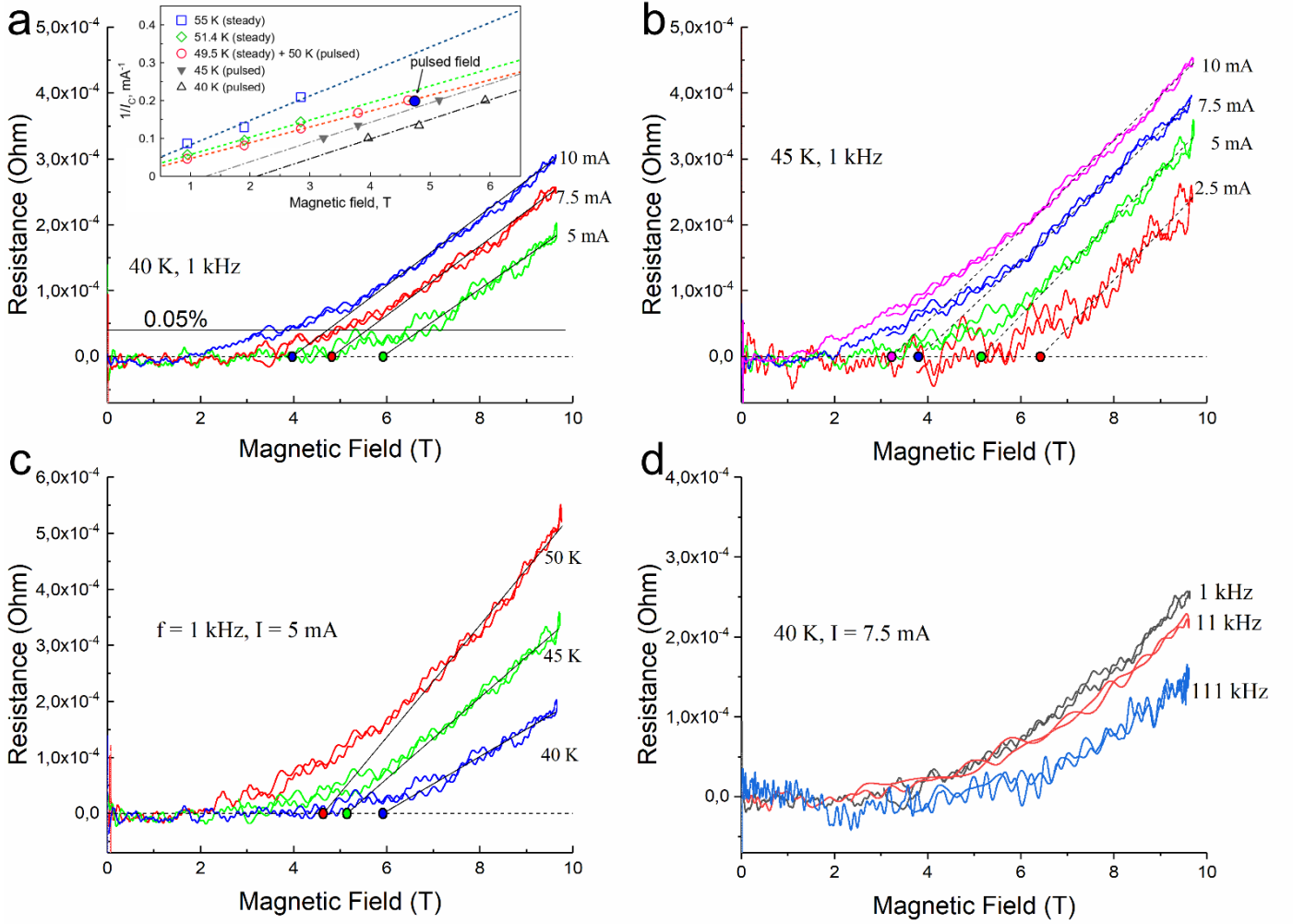

**Figure S7.** Critical magnetic fields at fixed AC currents in CeH$_9$ at temperatures 40-50 K. Resistance corresponds to the real impedance value (Re Z). (a) Transition to the non-superconducting state in CeH$_9$ in magnetic fields (*B*) of 0-10 T when various currents are applied at 40 K. AC current frequency was 1 kHz. Inset: a linear extrapolation of the *R(B)* in the normal state was used to determine the critical magnetic field at a fixed current. Inset: interpolation of the $I_C(B)$ data using the Bean-Kim model[16]. (b) Transition to the non-superconducting state in CeH$_9$ in magnetic fields of 0-10 T when various currents are applied at 45 K. (c) The same transitions in pulsed magnetic fields at 40, 45 and 50 K and fixed current *I* = 5 mA. (d) Comparison of transitions to the normal state in CeH$_9$ in a pulsed field at different measuring current frequencies with fixed amplitude (7.5 mA). Interestingly, but in case of high-frequency current (e.g., 111 kHz), the superconducting state in CeH$_9$ is preserved in stronger magnetic fields than in case of low-frequency current (1 kHz).

In general, the current-voltage characteristics of CeH$_9$ correspond to the behavior of other superconductors (for example, YH$_6$ [17], (La,Y)H$_{10}$ [4], SnH$_4$ [5]) and follows the Bean-Kim model

$$\frac{a}{I_C} = B + B_0, \qquad (1)$$

where $B_0 \approx 0.25$ T at 49.5 K [16], and below the critical current, the *V-I* form is a flat line $V \approx 0$, while when the critical current is exceeded, a sharp increase in resistance is observed. The values of critical currents measured at the SHMFF are given in Table S2, results of the HLD HZDR measurements are given in Table S3.

**Table S2.** Critical currents (in mA) of CeH$_9$ measured at different temperatures and steady magnetic fields (Figure S4) in AC mode. Due to sample heating, these critical currents must be considered as a lower estimate.

| Magnetic field, T | Temperature, K | | | |
|---|---|---|---|---|
| | 49.5 K | 51.4 K | 55 K | 69.6 K |
| 0 | - | - | - | 3.6 |
| 0.95 | 22.0 | 17.7 | 11.7 | |
| 1.9 | 12.5 | 10.6 | 7.8 | |
| 2.85 | 8.0 | 7.0 | 4.8 | |
| 3.8 | 6.0 | 4.0 | | |
| 4.75 | 5.0 | 1.7 | | |

**Table S3.** Critical magnetic fields (in Tesla) of CeH$_9$ measured at different temperatures and alternating currents (AC, 1 kHz). Measurements were done in pulsed magnetic fields at the HLD HZDR (Figure S5).

| Current, mA | Temperature, K | | |
|---|---|---|---|
| | 40 K | 45 K | 50 K |
| 5 | 5.92 | 5.15 | 4.64 |
| 7.5 | 4.82 | 3.8 | |
| 10 | 3.97 | 3.23 | |

## 3. Hall effect in CeH$_{9-10}$

The Hall effect studies were performed at the SHMFF by changing the DACs electrical connection: one of the voltage contacts was swapped with one of the current contacts. As a result, we obtained an almost linear odd dependence of the resistance $R_H(B)$ on the applied magnetic field (Figure S8). The measurements were carried out for a CeH$_{9-10}$ sample (DAC I1) in the non-superconducting state in the range from – 33.2 to + 33.2 T for two temperatures: 121.8 K and 129.6 K. In both cases, the Hall coefficient was almost the same. The sign of the Hall coefficient was not determined. Due to the asymmetric arrangement of the current and voltage contacts, there was an offset of 0.03 Ω when measuring the Hall resistance. This bias was corrected by subtracting $R_H(B=0) = 0.03$ Ω from the received $R_H(B)$ data. The measurements were carried out in AC mode with a current of 1 mA and a frequency of 13.333 Hz.

Experimentally found Hall coefficient, $R_H = 1/en_e t$, where $t$ – is the thickness of the sample, is in the range of $1.05\text{-}1.06\times10^{-4}$ Ω/T. If the sample thickness is assumed to be 1–3 μm, then the electron density $n_e$ in CeH$_{9-10}$ is $2\text{–}6\times10^{28}$ m$^{-3}$, which is compatible with the carrier concentration in H$_3$S:$(8.5 \pm 4.3)\times10^{28}$ m$^{-3}$ [18].

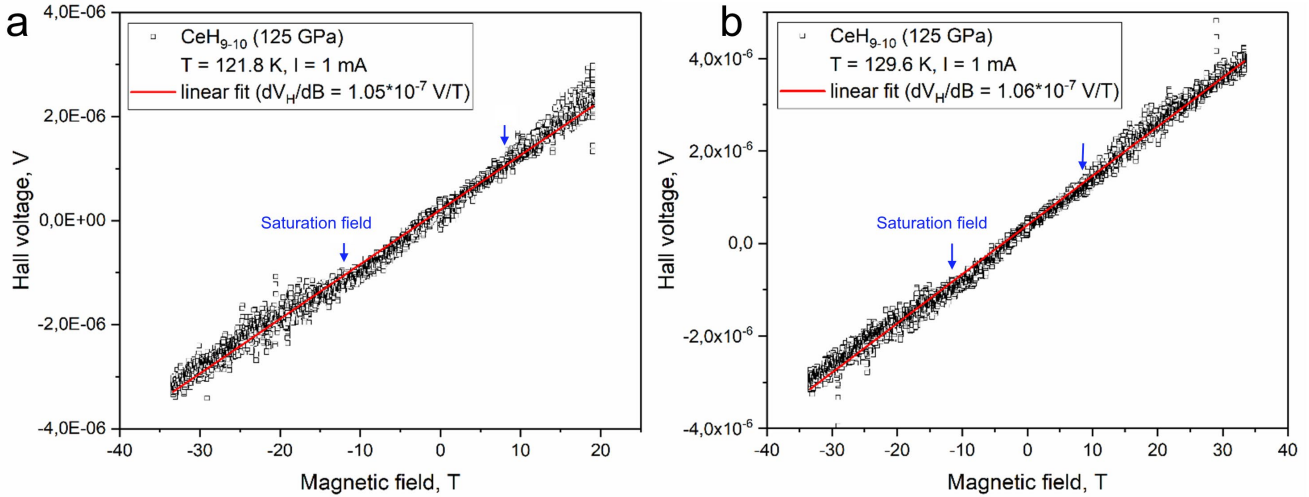

**Figure S8**. The Hall voltage on the DAC I1 sample versus applied magnetic field at (a) 121.8 K and (b) 129.6 K. A possible saturation field of magnetic impurities (or Ce atoms, see blue arrows) appears as an anomaly at 10-11 Tesla in the Hall voltage vs. temperature dependence.

The Hall coefficient of CeH$_9$ was also measured in DAC H1 at 95 K and 110 K (Figure S9). The results at these two temperature points were almost identical and for a sample thickness of 1–3 μm the Hall coefficient $R_H$ is about $3.6 – 10.8\times10^{-11}$ m$^3$/C. This value is slightly less than that obtained for the CeH$_{9-10}$ sample, and it leads to $n_e = 0.58 – 1.75\times10^{29}$ m$^{-3}$. Thus, the concentration of charge carriers (electrons) in cerium superhydrides is similar to other common metals, and corresponds to 1-3 electrons per unit cell.

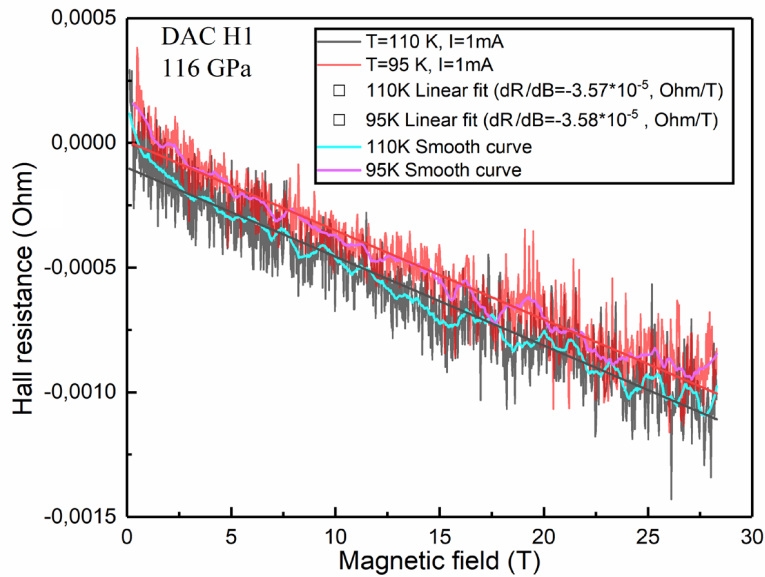

**Figure S9.** Hall resistance of the DAC H1 sample versus applied magnetic field at 95 K (red) and 110 K (grey).

## 4. Additional magnetoresistance and transport data

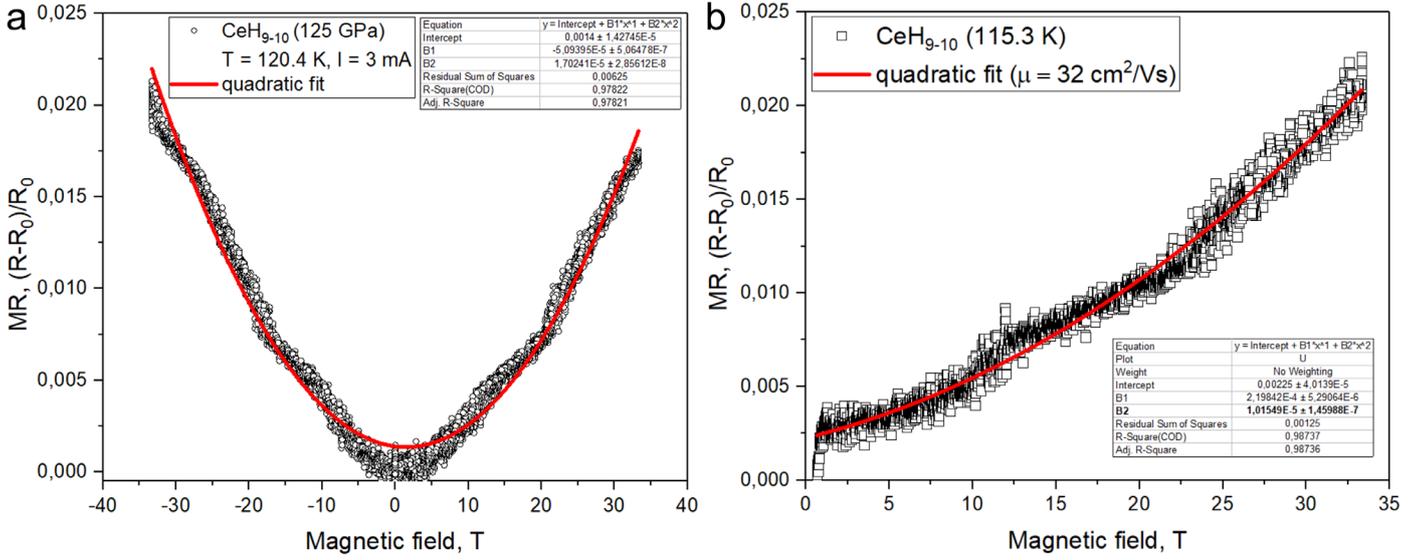

**Figure S10.** Quadratic ($\rho \propto \mu^2 B^2$) fit of experimental magnetoresistance of the DAC I1 sample at 120.4 K, above the critical temperature. The mobility of electrons ($\mu$) is 41 cm$^2$/s·V at 120.4 K, which is a typical value for metals.

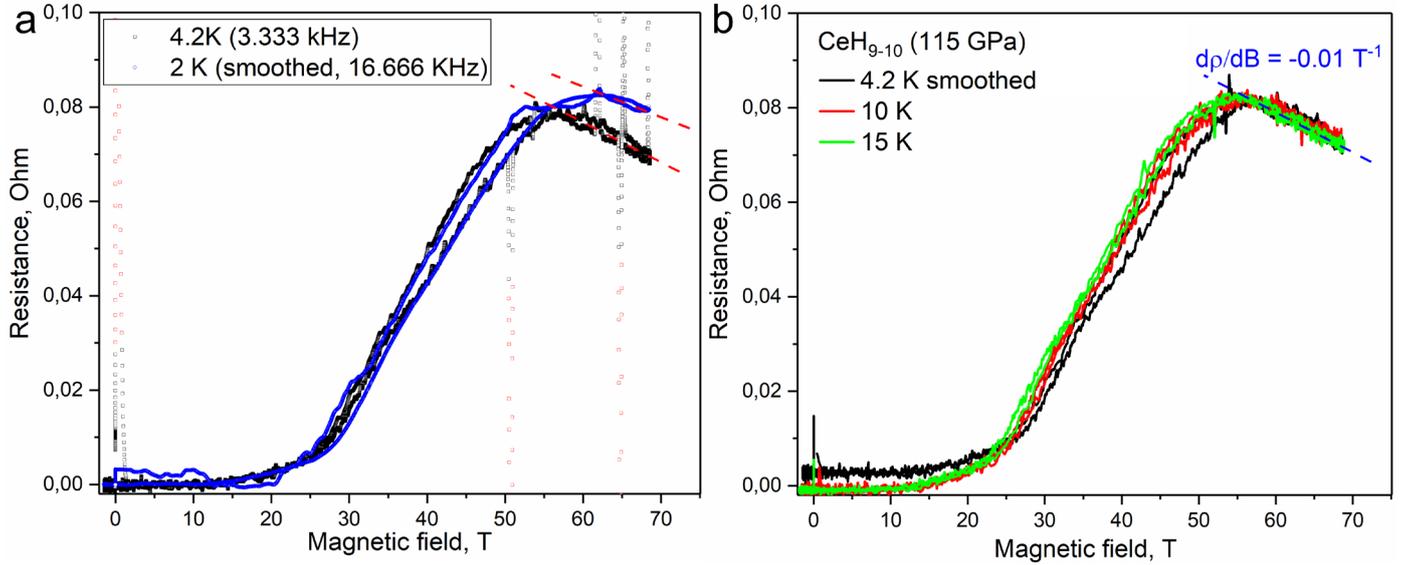

**Figure S11.** Low-temperature anomaly of $B_{C2}(T)$ in the CeH$_{9-10}$ (DAC I1) sample studied in pulsed fields at 115 GPa. (a) For temperatures of 2 K (blue) and 4.2 K (black). (b) For temperatures of 4.2–15 K. As can be seen, even at low temperatures there is a certain slope in the dependence $B_{C2}(T)$ and the $R(B)$ curves do not coincide leading to different $B_{C2}$. After the maximum resistance is reached, negative magnetoresistance (pseudogap phase) is observed and the electrical resistance decreases along with magnetic field induction. Moreover, it can be seen that the maximum resistance in the normal state at 2 K is significantly higher than at 4.2 K and is achieved in a higher magnetic field. The $R(B)$ curve at a temperature of 4.2 K was obtained at a measuring current frequency of 3.333 kHz. The remaining data were obtained at a frequency of 16.666 kHz.

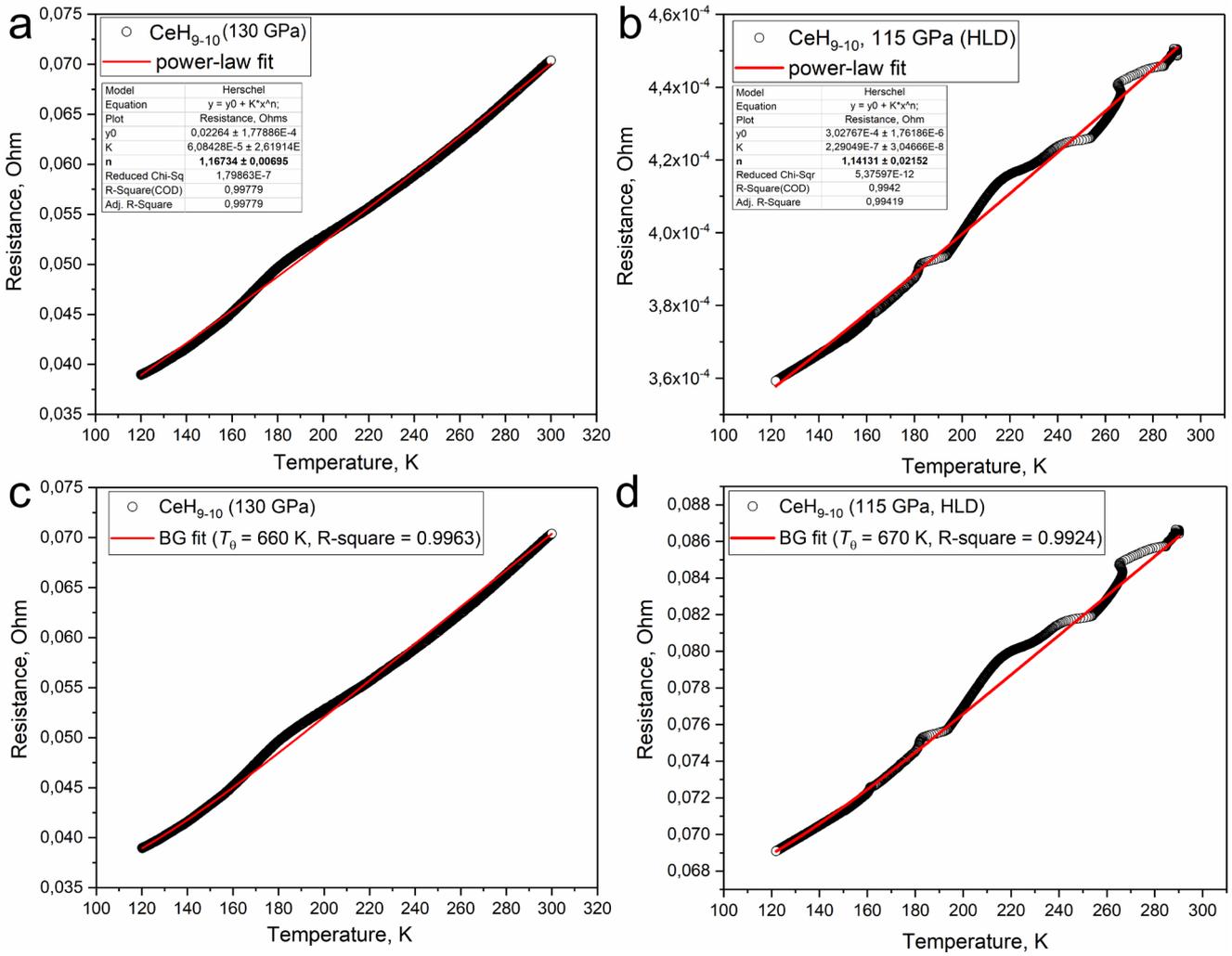

**Figure S12.** Dependence of electrical resistance of CeH$_{9-10}$ sample in DAC I1 on temperature in non-superconducting state and its various fits. (a, c) Power-law and the Bloch-Grüneisen fits of *R(T)* data obtained at LPI at pressure of 130 GPa, delta mode. (b, d) Power-law and the Bloch-Grüneisen fits of *R(T)* data obtained at HLD at pressure of 115 GPa, alternating current (AC) measurements. In both cases, the best interpolation of *R(T)* is given by a quasi-linear function.

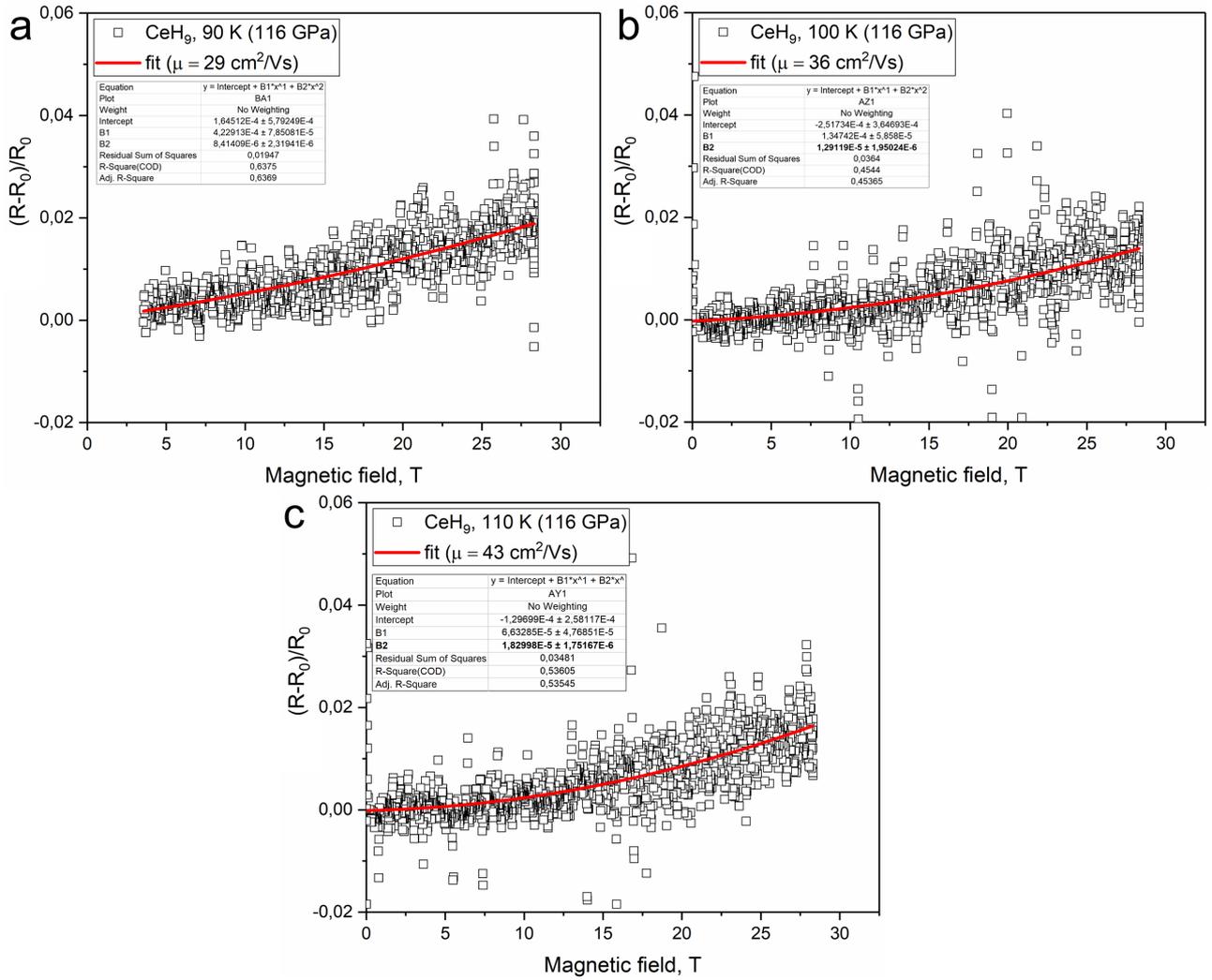

**Figure S13.** Quadratic ($\rho \propto \mu^2 B^2$) fit of experimental magnetoresistance of the DAC H1 sample at 90 (a), 100 (b) and 110 K (c), above the critical temperature. The mobility of electrons ($\mu$) is 29-43 cm$^2$/s·V, and decreases along with the temperature, as for the CeH$_{9-10}$ (DAC I1, Figure S10) sample.

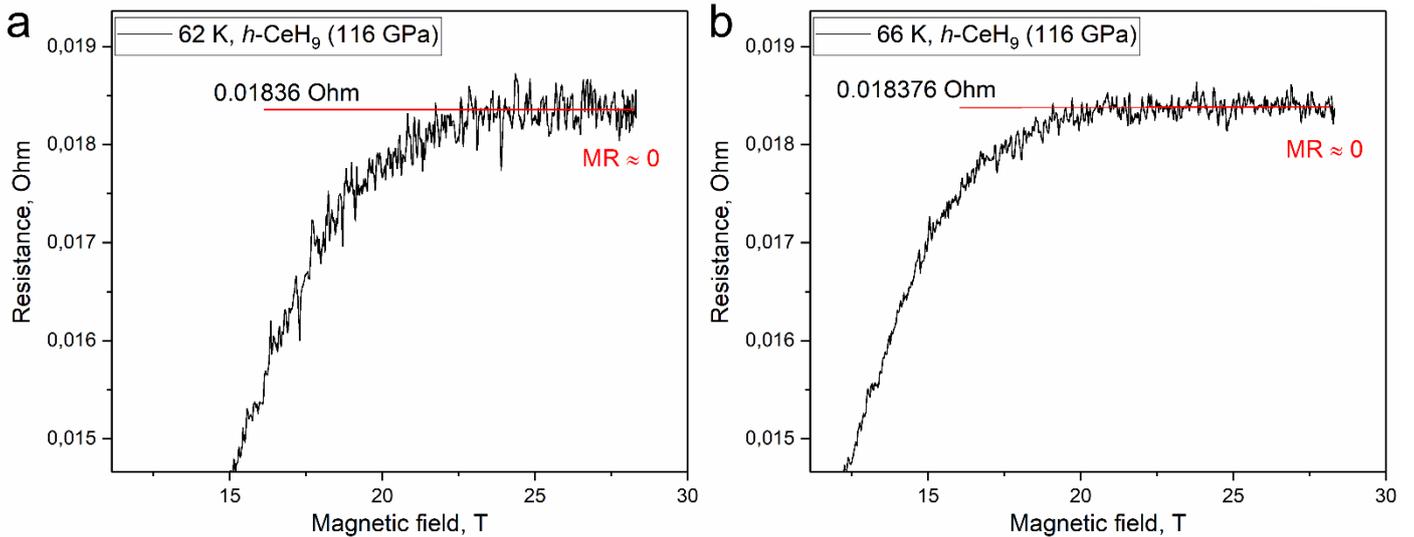

**Figure S14.** Dependence electrical resistance on the applied magnetic field of hexagonal (*h*) CeH$_9$ in DAC H1 at: (a) 62 K and (b) 66 K. Pressure is 116 GPa. At these temperatures, the magnetoresistance of the sample is almost zero.

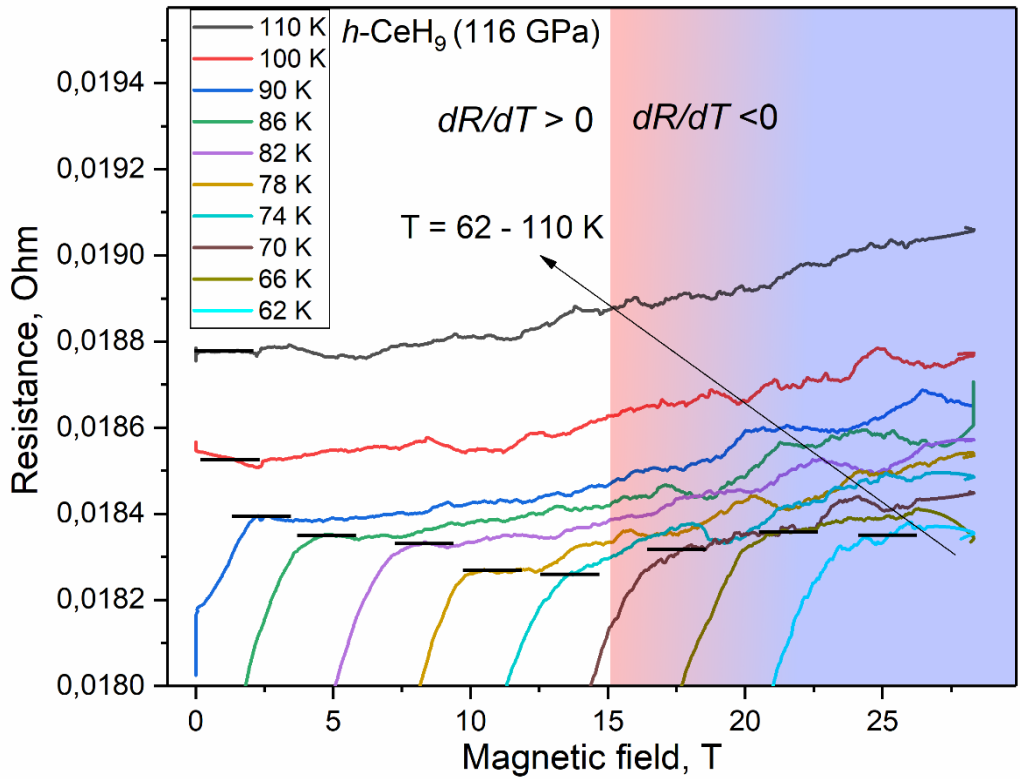

**Figure S15.** Dependence of the electrical resistance of CeH$_9$ (DAC H1) on the external magnetic field (0–27 T) for a series of temperature points from 62 K to 110 K. As the temperature decreases from 110 K to 74 K, there is a gradual decrease in the electrical resistance in the normal state right after $T_C$(onset), i.e. $dR/dT > 0$. It should be like that in common metals. Below 74 K, an anomalous increase in resistance is observed along with decreasing temperature ($dR/dT < 0$).

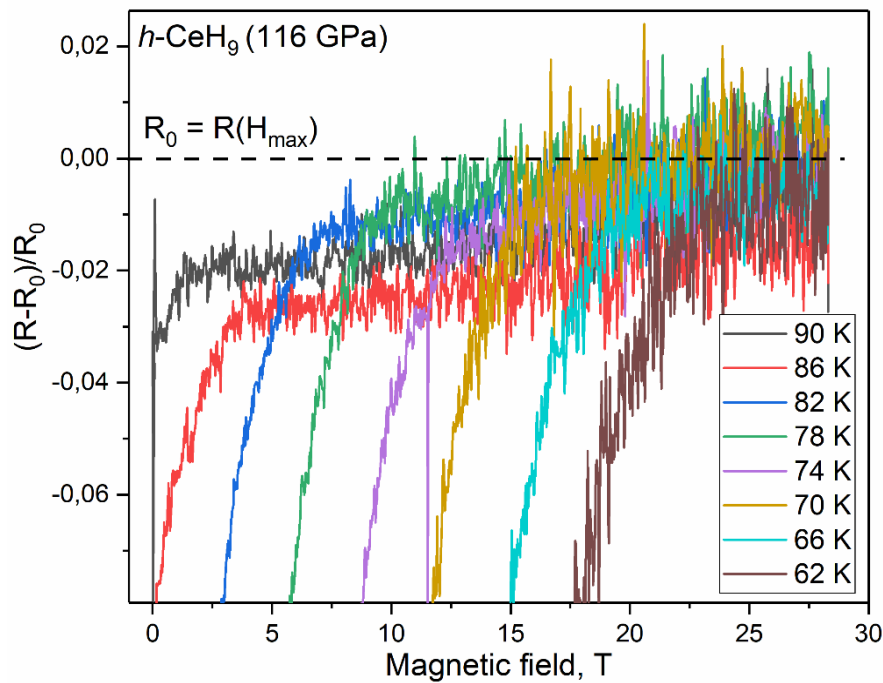

**Figure S16.** Dependence of the reduced resistance $\rho = (R - R_0)/R_0$ of a sample of hexagonal cerium hydride $h$-CeH$_9$ (DAC H1) at 116 GPa on an external magnetic field of 0 – 27 T at different temperatures. $R_0$ is taken as the electrical resistance in the non-superconducting state in the maximum magnetic field for each temperature point.

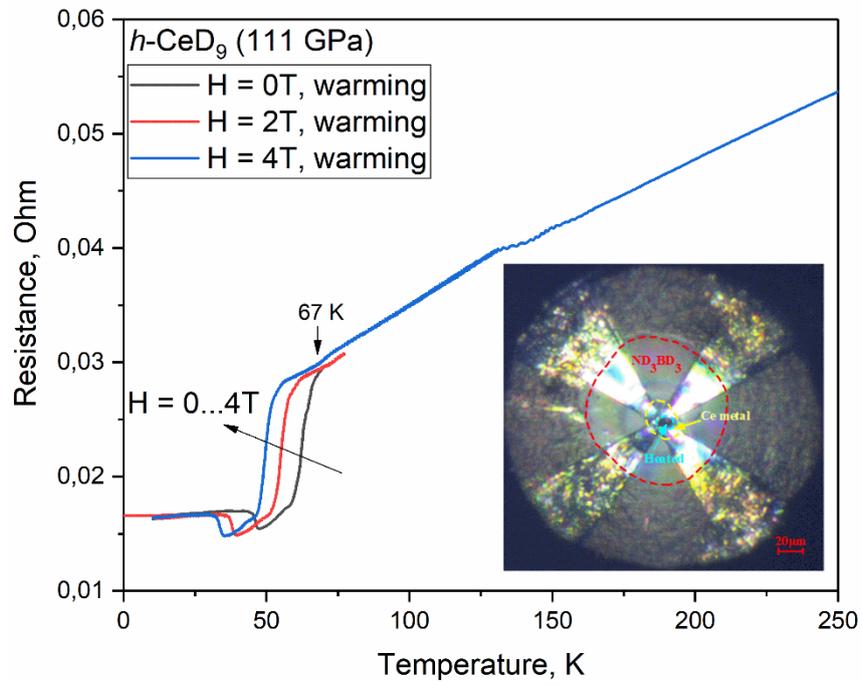

**Figure S17**. Temperature dependence of the electrical resistance of hexagonal (*h*) cerium deuteride *h*-CeD$_9$ (DAC D1, warming cycle) in the presence of an external magnetic field of 0, 2, and 4 Tesla at 111 GPa. The drop in resistance at 67 K corresponds to the superconducting transition in CeD$_9$ described in Ref. [19]. Inset: photo of the DAC D1 sample, deuterated ammonia borane ND$_3$BD$_3$ and 4-electrode van der Pauw circuit for transport measurements.

## 5. X-ray diffraction data

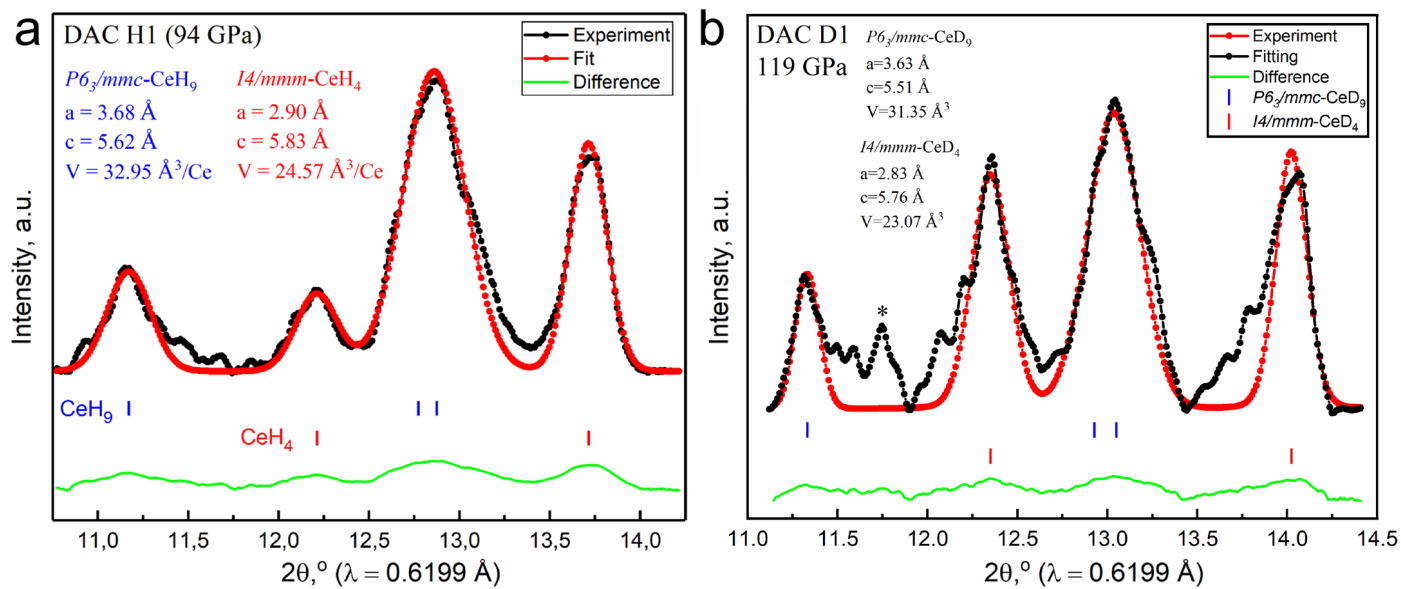

**Figure S18**. X-ray diffraction pattern of studied samples and the Le Bail refinements of the unit cells parameters. (a) DAC H1 at 94 GPa. Le Bail refinement of the CeH$_9$ and CeH$_4$ unit cells. (b) DAC D1 at 119 GPa. Le Bail refinement of the CeD$_9$ and CeD$_4$ unit cells.

# 6. Theoretical calculations

Analysis of the temperature dependence of electrical resistance (Figure 1) using the Bloch-Grüneisen [7-8] and Allen-Dynes (AD) [20] formulas shows that in the case of $CeH_{9-10}$ (DAC I1) sample the EPC coefficient at 130 GPa is unexpectedly high ($\lambda \approx 2.5$) compared to our previous results [19] which calls into question the applicability of this interpolation. A retrospective analysis of the data that we obtained in 2021 (Figure S19, Table S3) indicates a low reproducibility of the electron-phonon interaction parameters obtained using Bloch-Grüneisen (BG) fit or indicates an anomalous pseudo-linear dependence of the electrical resistance on temperature $R(T)$, which leads to an underestimation of the Debye temperature and an overestimation of the electron-phonon interaction strength. For example, based on all experiments, we can only give a rough estimate for the electron-phonon interaction strength $\lambda = 2.0 \pm 0.5$ in $CeH_{9-10}$ at 110-130 GPa (Figure S19d).

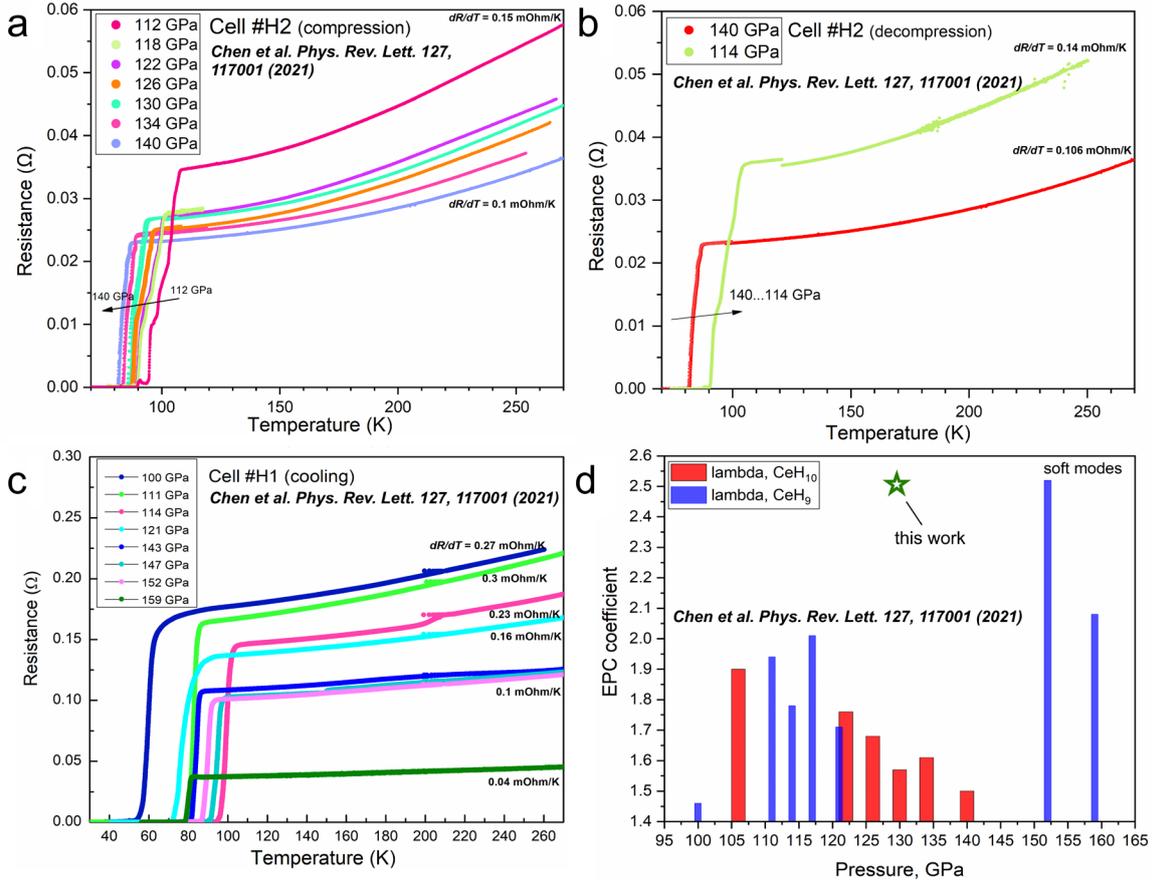

**Figure S19.** Temperature dependence of electrical resistance R(T) for various cerium hydride samples taken from Ref.[19]. (a, b) DAC H2, main component is $Fm\bar{3}m$-$CeH_{10}$. (c) DAC H1, main component is $P6_3/mmc$-$CeH_9$. (d) Pressure dependence of the electron-phonon interaction strength ($\lambda$) obtained from $R(T)$ and $T_C$ for $CeH_9$ and $CeH_{10}$.

Anisotropy of the superconducting gap in the hexagonal $CeH_9$ and the electrical resistivity were studied using Quantum ESPRESSO [10-11] and EPW codes [21-22] at pressure of 100 GPa in harmonic approximation. We used 8×8×8 k-mesh and 4×4×4 q-mesh, upper limit over frequency integration/summation in the Eliashberg equations was 6 eV, convergence threshold on imaginary axis was $10^{-4}$, lowest boundary for the phonon frequency was taken 100 cm$^{-1}$, smearing in the energy-conserving delta functions was $\sigma = 0.05$–0.5 eV. Obtained density of electron states of $hcp$ $P6_3/mmc$-$CeH_9$ at 100 GPa was 0.718 states/eV/Ce atom. Analysis (Figure S20) shows that superconductivity in $CeH_9$ has anisotropic character and $\Delta_{nk}(0) = 31$–39 meV. However, solving the anisotropic Eliashberg equations using the EPW code leads to a significantly overestimated $T_C$(EPW) ≈ 192 K at 100 GPa. In general, this short computational study confirms that the anisotropy of the electron-phonon interaction may be responsible for high experimental values of $T_C > 110$ K exceeding the results of calculations within the isotropic approach ($\approx 80$–100 K).

The overestimation of $T_C$ is typical for many anisotropic calculations of superconductivity in polyhydrides. Calculations show that within the isotropic approximation (interpolation method [14-15]) the obtained electron-phonon coupling parameters are higher than found by the tetrahedron method [19]: $\lambda$= 1.98, $\omega_{log}$ = 712 K, and the critical temperature $T_C$(AD) = 102 K approaches the experimental values.

**Table S3.** The parameters of the electron-phonon interaction in the samples of CeH$_9$ and CeH$_{10}$ found from the retrospective analysis of temperature dependence of electrical resistance $R(T)$, taken from Ref. [19], using BG and Allen-Dynes formulas. For simplicity we used $\mu^*$=0.1, $\omega_2$ = (4/3)×$\omega_{log}$, $\omega_{log}$ = 0.827×$T_D$.

| Pressure, GPa | $T_D$, K | $\omega_{log}$, K | $\lambda$ | $T_C$, K | dR/dT, mOhms/K |
|---|---|---|---|---|---|
| $Fm\bar{3}m$-CeH$_{10}$ | | | | | |
| 106 | 785 | 650 | 1.9 | 108 | 0.15 |
| 122 | 790 | 653 | 1.76 | 101 | 0.125 |
| 126 | 790 | 653 | 1.68 | 96 | 0.12 |
| 130 | 830 | 686 | 1.57 | 94 | 0.12 |
| 134 | 775 | 641 | 1.61 | 90 | 0.1 |
| 140 | 800 | 660 | 1.5 | 86 | 0.1 |
| $P6_3/mmc$-CeH$_9$ | | | | | |
| 100 | 605 | 500 | 1.46 | 64 | 0.27 |
| 111 | 620 | 513 | 1.94 | 87 | 0.3 |
| 114 | 805 | 665 | 1.78 | 104 | 0.23 |
| 117 | 655 | 541 | 2.01 | 95 | 0.28 |
| 121 | 670 | 554 | 1.71 | 83 | 0.16 |
| 152 | 530 | 438 | 2.52 | 94 | 0.1 |
| 159 | 540 | 446 | 2.08 | 81 | 0.04 |

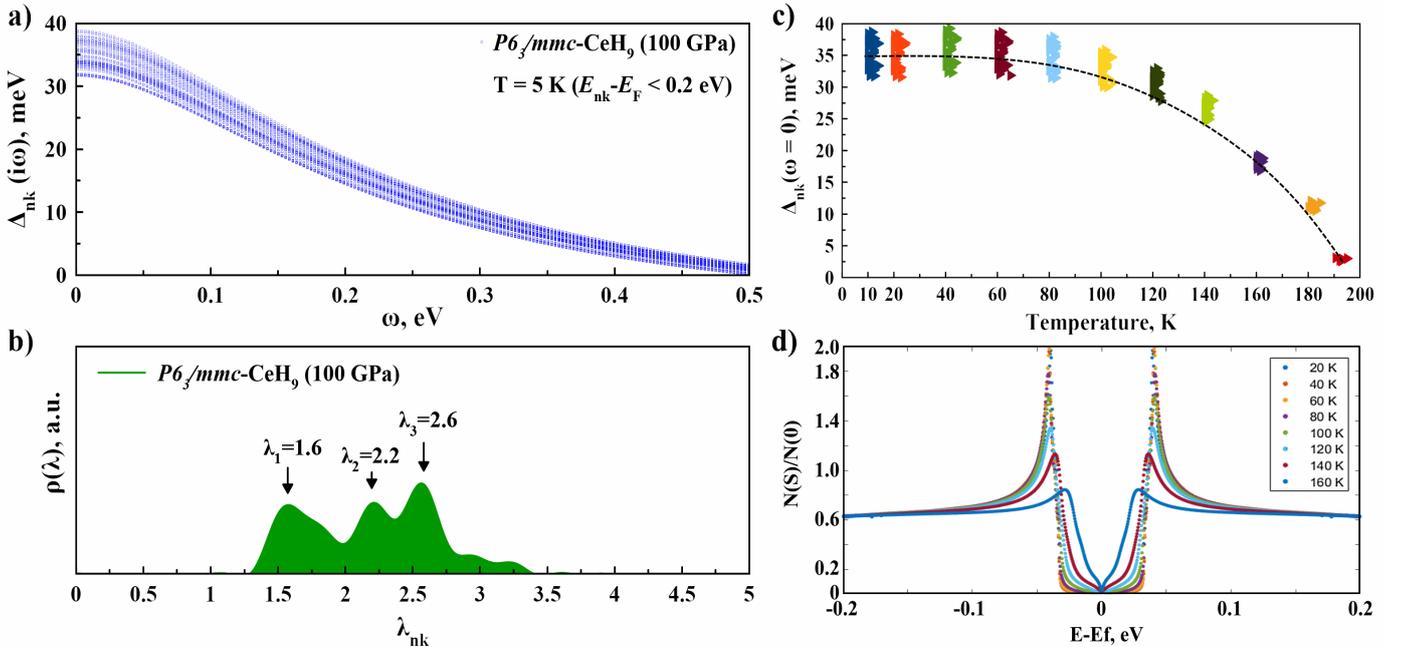

**Figure S20.** Results of anisotropic calculations for CeH$_9$. (a) Calculated energy-dependent superconducting gap along the imaginary energy axis of hcp-CeH$_9$ at 5 K and 100 GPa. Only the Kohn-Sham states (nk) with an energy within 0.2 eV from the Fermi level are shown. (b) Distribution of k-resolved EPC constants $\lambda_{nk}$. (c) Calculated anisotropic superconducting gap $\Delta_{nk}(\omega=0)$ of hcp-CeH$_9$ on the Fermi surface, evaluated as a function of temperature. The Coulomb pseudopotential is $\mu_* = 0.1$. For each temperature the histograms indicate the number of states on the Fermi surface with that superconducting gap energy. The dashed lines are guides to the eye. The superconducting gap vanishes at the critical temperature $T_C \approx$ 192 K, which is much higher than the experimental $T_C \approx$ 110-117 K. (d) Quasiparticle DOS in the SC state for various temperatures obtained using $\Delta(\omega=0)$.

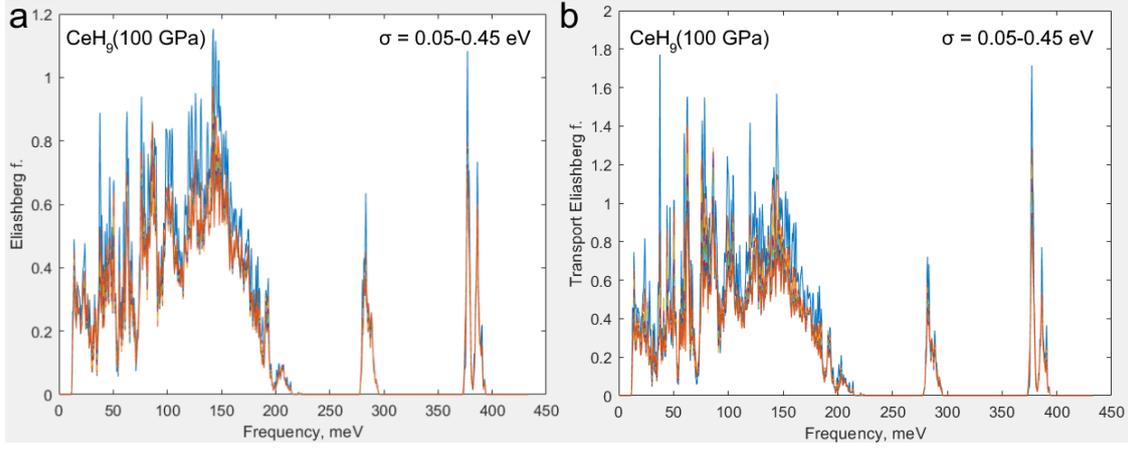

**Figure S21.** Calculated isotropic Eliashberg (a) and transport Eliashberg (b) spectral functions of *hcp*-CeH$_9$ at 100 GPa and different gaussian smearing values ($\sigma$ = 0.05-0.45 eV) used for calculations in the Quantum Espresso.

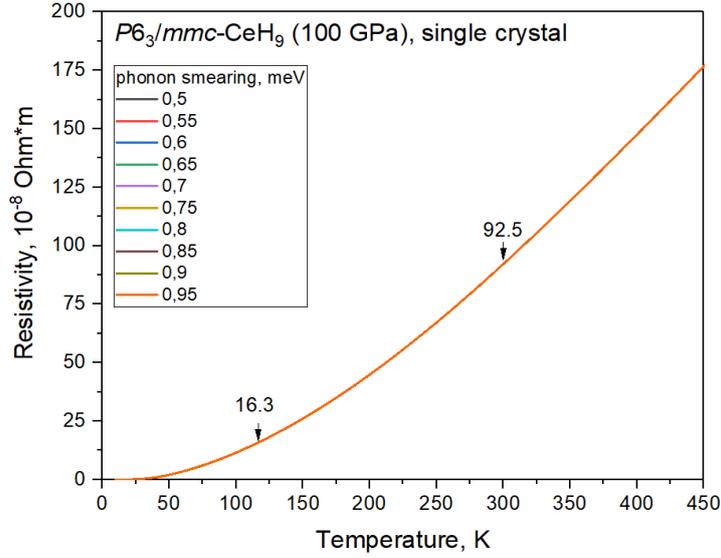

**Figure S22.** Calculated resistivity of the CeH$_9$ at 100 GPa as a function of temperature calculated using the Eliashberg transport function given in Figure S19b. Values shown by arrows correspond to resistivity at 300 K and at the critical temperature of 117 K at 130 GPa.

The calculated resistivity (Figure S22) can be used to estimate the geometric dimensions of the CeH$_{9-10}$ samples, in particular, its thickness. We assume that the resistivity of the sample is the sum of the scattering on phonons $\rho_s(T)$ and scattering on defects ($\rho_0$): $\rho_{full} = \rho_s(T) + \rho_0$, then, by definition, for rectangular sample

$$R = \rho_{full} \times \left(\frac{d}{S}\right), \quad (S2)$$

where $d$ – is the sample size, and $S$ – is the cross-section area). Using the formula for the symmetric Van der Pauw circuit [23] we can write

$$R = \frac{\rho_{full}}{\pi t} \ln 2. \quad (S3)$$

Now, using the data on the resistance $R(T)$ (Figure 1a), we can find all the parameters of interest, e.g.

$$\rho_0 = \frac{R(T_2)\rho_s(T_1) - R(T_1)\rho_s(T_2)}{R(T_1) - R(T_2)}. \quad (S4)$$

In our case, $(t/S) \approx 1/d$, where $t$ – is the thickness. Using the data from Figures 1 and S22, we get $d$ = 25 μm for DAC I1, which is close to the actual sample size. If we use the formula for the Van der Pauw circuit (eq. S3), then we get $t$ = 5.4 μm, which is also reasonable, although somewhat higher than the interval used in the main text (1–3 μm). The resistance of the CeH$_9$ sample in DAC H1 is approximately two times lower (18 mΩ), which corresponds to its greater thickness.